\documentclass{article}
\usepackage[english]{babel}
\usepackage[utf8]{inputenc}
\usepackage{amsmath}
\usepackage{amssymb}
\usepackage{siunitx}
\usepackage{xcolor}
\usepackage[colorinlistoftodos]{todonotes}
\usepackage{authblk}
\usepackage{lmodern}
\usepackage{url}

\usepackage{IEEEtrantools}

\usepackage{bm} 

\usepackage[margin=2cm]{geometry}
\usepackage{calc}

\usepackage[ruled,vlined]{algorithm2e} 

\usepackage{graphicx}
\graphicspath{{Images/}}
\usepackage{standalone}
\usepackage{multirow}
\usepackage{morefloats}
\usepackage{caption}
\usepackage{subcaption}
\usepackage{epstopdf}

\usepackage[toc,page]{appendix}
\usepackage[numbers]{natbib}

\usepackage{tikz}
\usepackage{tikz-3dplot}
\usepackage{pgfplots}
\usetikzlibrary{hobby}
\pgfplotsset{compat=1.15}
\usepackage[multiple]{footmisc}
\usetikzlibrary{calc}
\usetikzlibrary{decorations.pathmorphing, patterns,shapes}

\pgfmathdeclarefunction{genTri}{3}{%
  \pgfmathparse{and(#1 <= x, x <= #3)*(2*(x-#1)/((#2-#1)*(#3-#1))) + and(#3 < x, x <= #2)*(2*(#2 - x)/((#2-#1)*(#2-#3)))}%
}

\pgfmathdeclarefunction{gauss}{2}{%
  \pgfmathparse{1/(#2*sqrt(2*pi))*exp(-((x-#1)^2)/(2*#2^2))}%
}

\pgfmathdeclarefunction{symTri}{2}{%
  \pgfmathparse{and(#1<=x, x <= ((#1+#2)/2)) * (4*(x-#1)/(#2-#1)^2) + 
    and(((#1+#2)/2) < x, x <= #2) * (4*(#2-x)/(#2-#1)^2)
  }%
}

\usepackage{verbatim}

\makeatletter

\newcommand{\Rmnum}[1]{\expandafter\@slowromancap\romannumeral #1@}
\makeatother

\title{Bayesian inference on a microstructural, hyperelastic model of tendon deformation}
\author[1]{James Haughton}
\author[1]{Simon L. Cotter}
\author[1]{William J. Parnell}
\author[1, 2]{Tom Shearer}
\affil[1]{Department of Mathematics, University of Manchester, Manchester M13 9PL, United Kingdom}
\affil[2]{Department of Materials, University of Manchester, Manchester M13 9PL, United Kingdom}
\date{}

\begin{document}

\maketitle

\begin{abstract}
Microstructural models of soft tissue deformation are important in applications including artificial tissue design and surgical planning. The basis of these models, and their advantage over their phenomenological counterparts, is that they incorporate parameters that are directly linked to the tissue's microscale structure and constitutive behaviour and can therefore be used to predict the effects of structural changes to the tissue. Although studies have attempted to determine such parameters using diverse, state-of-the-art, experimental techniques, values ranging over several orders of magnitude have been reported, leading to uncertainty in the true parameter values and creating a need for models that can handle such uncertainty. We derive a new microstructural, hyperelastic model for transversely isotropic soft tissues and use it to model the mechanical behaviour of tendons. To account for parameter uncertainty, we employ a Bayesian approach and apply an adaptive Markov chain Monte Carlo algorithm to determine posterior probability distributions for the model parameters. The obtained posterior distributions are consistent with parameter measurements previously reported and enable us to quantify the uncertainty in their values for each tendon sample that was modelled. This approach could serve as a prototype for quantifying parameter uncertainty in other soft tissues.
\end{abstract}


\section{Introduction}

\begin{figure}
   \begin{subfigure}[t]{.485\textwidth}
    \centering
    \includegraphics{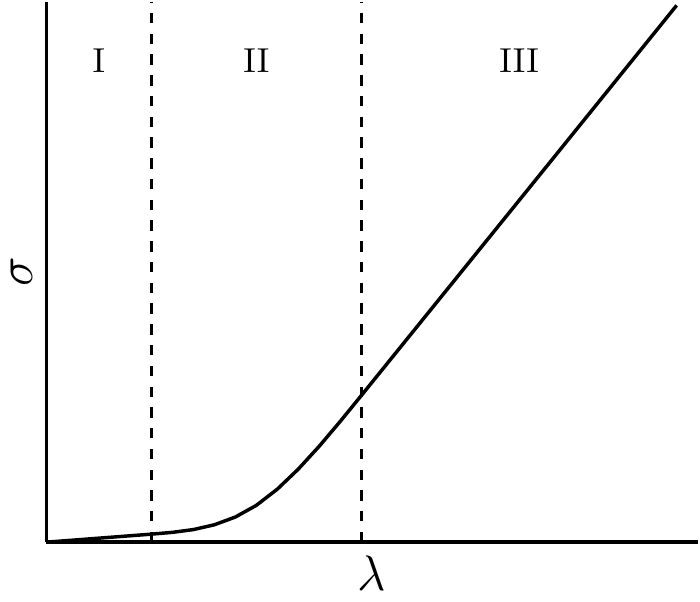}
    \subcaption[]{}
    \label{subfig:stress_strain_behaviour}
  \end{subfigure}
  \hfill
  \begin{subfigure}[t]{.485\textwidth}
    \centering
    \includegraphics{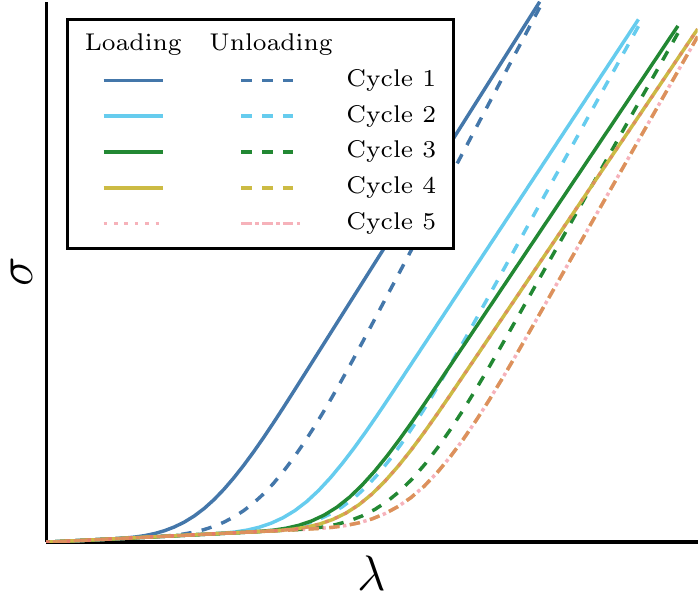}
    \subcaption[]{}
    \label{subfig:pseudoelasticity}
  \end{subfigure}
    \caption{(a) The stress-strain behaviour of soft tissues. Region \Rmnum{1}: only the compliant components are loaded; the collagen fibrils are crimped and slack. Region \Rmnum{2}: gradually the stiff collagen fibrils straighten and become taut. Region \Rmnum{3}: all the collagen fibrils are taut; the soft tissue is stiff and linearly elastic. (b) Successive loading-unloading cycles of a viscoelastic soft tissue until the tissue can be treated as pseudoelastic.}
    \label{fig:softTissueBehaviour}
\end{figure}

Fibrous soft tissues such as tendons, skin, and arteries are vital to life. Tendons and ligaments, for example, enable movement by transmitting forces around the body \cite{james2008tendon}. It is critical, therefore, that we understand soft tissue mechanical behaviour to advance fields such as tissue engineering \cite{geris2018future} and surgery \cite{famaey2008soft}. Soft tissues exhibit complex macroscopic phenomena, including anisotropy and non-linearity, that are induced predominantly by the microstructure of the tissue. Anisotropy arises from the presence of collagen fibrils, which locally reinforce the tissue in a preferred direction. Initially, the fibrils are crimped and stress-free, but they straighten as the tissue deforms, contributing to its resistance to further deformation once taut \cite{franchi2007crimp}. This gradual recruitment of collagen fibrils leads to the non-linear stress-strain profile typical of soft tissues \cite{hamdia2019sensitivity}, as illustrated in Figure \ref{subfig:stress_strain_behaviour} with a plot of the Cauchy stress, $\sigma$, against stretch, $\lambda$. 

Additionally, soft tissues are \textit{viscoelastic}, so assuming that their behaviour can be described by an elastic model is a simplification. Practically speaking, before tests to measure mechanical properties are performed, a tissue is subjected to cyclic loading until the stress-strain behaviour of the tissue is consistent between consecutive cycles (see Figure \ref{subfig:pseudoelasticity}). Then, the tissue can be treated as pseudoelastic and modelled as a particular elastic material upon loading and a different elastic material upon unloading \cite{fung1980pseudo}. In reality, energy is dissipated in the tissue during the loading-unloading cycle, but we can apply elasticity theory to the tissue as long as we only examine one loading path. Furthermore, for sufficiently slow (quasi-static) or extremely rapid deformations, the loading and unloading curves are relatively close to one another as hysteresis is minimised in these regimes.

To model soft tissue deformation, we will use the theory of hyperelasticity, relating the stress to the strain via a strain-energy function (SEF). There are two approaches to developing a hyperelastic model: the phenomenological and structural approaches (although any one SEF can incorporate features of both). Phenomenological models seek to achieve the best quantitative fit to experimental data. They do not attempt to determine how the microstructure influences the macroscopic behaviour observed in mechanical testing because the model's parameters do not necessarily have a clear physical interpretation. By contrast, structural models incorporate physically relevant parameters to elucidate the relationship between the arrangement and properties of the tissue's constituents and its mechanical behaviour. Incorporating microstructural information into an SEF often increases its complexity and it is important that a structural SEF remains tractable if it is to be viable for studying soft tissue deformation. This is important when employing a Bayesian framework with Monte Carlo sampling, where we require a large number of solutions of the forward model. Therefore, simplifying assumptions about the microstructure are often required.

Values for unknown structural parameters can be obtained via imaging methods such as serial block face-scanning electron microscopy \cite{starborg2013using,chang2020circadian} and X-ray computed tomography \cite{shearer2014x,shearer2016three,balint2016optimal,rawson2018four}, and for constitutive parameters using micromechanical techniques like force spectroscopy \cite{graham2004structural} and atomic force microscopy \cite{svensson2012mechanical}. These techniques are challenging, however, and a wide range of values has been reported for certain quantities. The collagen fibril Young's modulus, for example, has been reported to have a value ranging from 32 MPa \cite{graham2004structural} to 2.8 GPa \cite{svensson2012mechanical}. This uncertainty makes it difficult to predict soft tissue mechanical behaviour using optimisation techniques alone. Therefore, in this paper we take a Bayesian approach to the modelling process to characterise the likely ranges of values that microstructural and micromechanical parameters can take. 

Due to their importance and the fact that they have been studied extensively, we focus on tendons in this paper. The mechanical properties of different tendons are distinct from one another, with energy-storing tendons being more extensible than positional tendons due to differences in their microstructures \cite{thorpe2012specialization,shearer2017relative}. One feature that is common to all tendons is that their collagen is structured in a regulated, hierarchical fashion and aligned closely with the tendon's axis \cite{james2008tendon}. Collagen molecules form cross-links and aggregate into fibrils with diameters ranging from 12 to 500 nm \cite{starborg2013using}. Collections of fibrils collect into larger structures called fibres, with diameters of 150 to 1000 $\mu$m, which themselves form fascicles, with diameters of 1000 to 3000 $\mu$m, \cite{kannus2000structure}. In other soft tissues, collagen fibrils are less strongly aligned and form a network, but by aligning many fibrils in one direction, the tendon is stronger in that direction \cite{revell2021collagen}. 

Models such as the Holzapfel-Gasser-Ogden (HGO) model \cite{holzapfel2001new}, which was initially created to study arteries, have been adapted to study tendons \cite{akintunde2018evaluation}. This model is structural in the sense that it incorporates a strain invariant that is directly related to the stretch in the collagen fibres, but phenomenological in the sense that an exponential function is used to describe collagen recruitment, and the stretches in individual fibrils are not tracked. Other models have explicitly incorporated the crimp morphology of the fibrils \cite{freed2016promising, shearer2015new, shearer2015helical, shearer2020recruitment,gregory2021microstructural} and produced a good fit to experimental data. Several probability density functions (PDFs) have been used to describe the distribution of fibril length, including the Weibull \cite{hurschler1997structurally} and triangular distributions \cite{watton2004mathematical,aparicio2016novel,bevan2018biomechanical}, as summarised in a review article by Thompson \textit{et al.} \cite{thompson2017mechanobiological}.

In this paper, we derive a new structural SEF for modelling soft tissues that assumes collagen fibrils are linearly elastic and have a triangular length distribution. We test the efficacy of the model using non-linear optimisation to find a parameter vector that produces a local best fit to data. Secondly, we repeat the fitting process using a Bayesian framework. This enables us to incorporate prior beliefs about the unknown model parameters, a statistical model for noisy observations of the stress-strain curves and our non-linear model to obtain posterior distributions for the model parameters. Through these distributions, we identify and quantify the uncertainty in the parameters, and the directions in parameter space in which the model is more or less sensitive.

The structure of the paper is as follows. In Section 2, we describe the underpinning continuum mechanical theory that is required to model the deformation of an anisotropic soft tissue and, using physical considerations, derive a new constitutive equation to model tendons. In Section 3, we use non-linear optimisation to fit the model to experimental stress-strain data and compare its quality of fit to that of the widely-used HGO model and a microstructural tendon model. In Section 4, we account for noise in the experimental data using a likelihood function that allows us to study the problem under a Bayesian framework. In Section 5, we derive the posterior distribution for the model's microstructural and micromechanical parameters. In Section 6, we summarise our findings and discuss potential ways to expand upon our work.

\section{Model derivation}

\subsection{Preliminaries}

Prior to considering the constitutive response of soft tissues, we need to consider kinematics, i.e.\ how to formulate the mechanism of deformation. First, we distinguish between two configurations, the reference (initial) configuration and the deformed configuration. Points on the reference and deformed bodies are described by the vectors $\mathbf{X}$ and $\mathbf{x}$, respectively. The two sets of coordinates are related via the deformation mapping, $\chi$, i.e.\ $\mathbf{x} = \chi(\mathbf{X})$. We define the deformation gradient, $\mathbf{F}$, as

\begin{equation}
    \label{eq:deformGradient}
    \mathbf{F} = \nabla_\mathbf{X} \mathbf{x},
\end{equation}
where $\nabla_\mathbf{X}$ represents the gradient operator with respect to the reference coordinates. From the deformation gradient, we define two symmetric measures of the deformation, known as the left and right Cauchy-Green deformation tensors, $\mathbf{B}= \mathbf{F}\mathbf{F}^\textrm{T}$ and $\mathbf{C}= \mathbf{F}^\textrm{T}\mathbf{F}$, respectively \cite{holzapfel2000nonlinearbook}.

The SEF $W$ allows one to define the constitutive equation of a hyperelastic material, relating stress to strain via derivatives of $W$. In order to determine the exact form of this constitutive response, we must first identify the symmetry properties of the material. The mechanical behaviour of transversely isotropic materials, such as tendons, is only invariant for rotations around a preferred direction, $\mathbf{M}$. Furthermore, the SEF must be objective as the laws of physics are the same in any inertial frame of reference. As the SEF is invariant under a coordinate transformation, we can write it as a function of invariants of the deformation. For an isotropic material, there are only three invariants, $I_1=\text{tr}(\mathbf{C})$, $I_2=\frac{1}{2}((\text{tr}(\mathbf{C}))^2-\text{tr}(\mathbf{C}^2))$ and $I_3=\det(\mathbf{C})$. For a transversely isotropic material, we must introduce an additional two pseudoinvariants that depend on $\mathbf{M}$: $I_4=\mathbf{M}\cdot\mathbf{CM}$ and $I_5=\mathbf{M}\cdot\mathbf{C}^2\mathbf{M}$ \cite{holzapfel2000nonlinearbook}.

\subsection{The model} \label{sec:themodel}

\begin{figure}
  \begin{subfigure}[t]{.485\textwidth}
    \centering
    \includegraphics{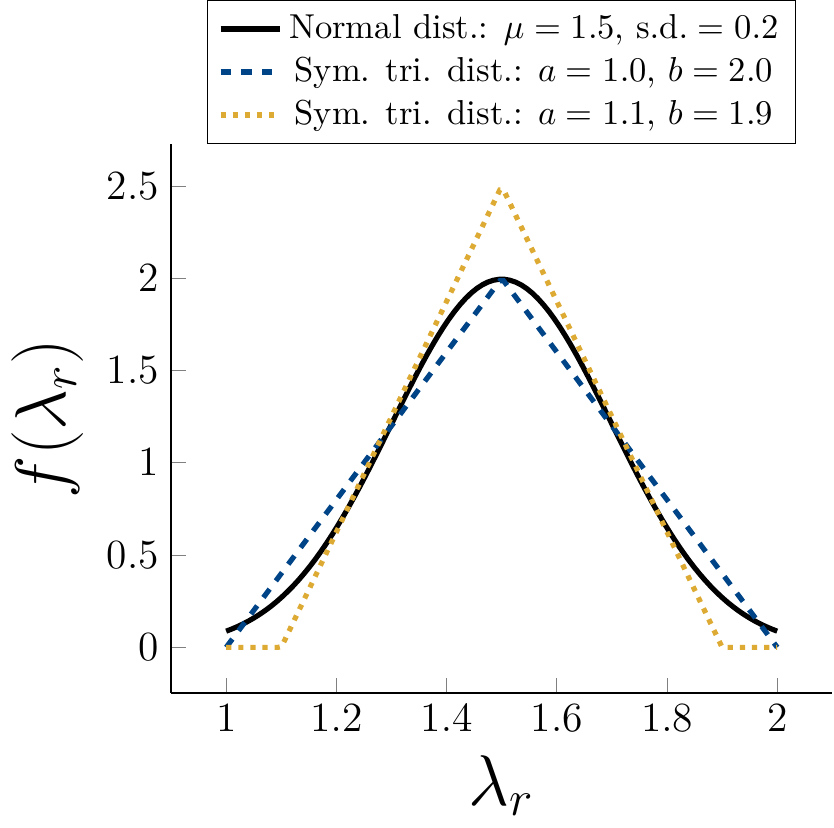}
    \subcaption[]{}
    \label{subfig:normalAndSymTriDists}
  \end{subfigure}
  \hfill
  \begin{subfigure}[t]{.485\textwidth}
    \centering
    \includegraphics{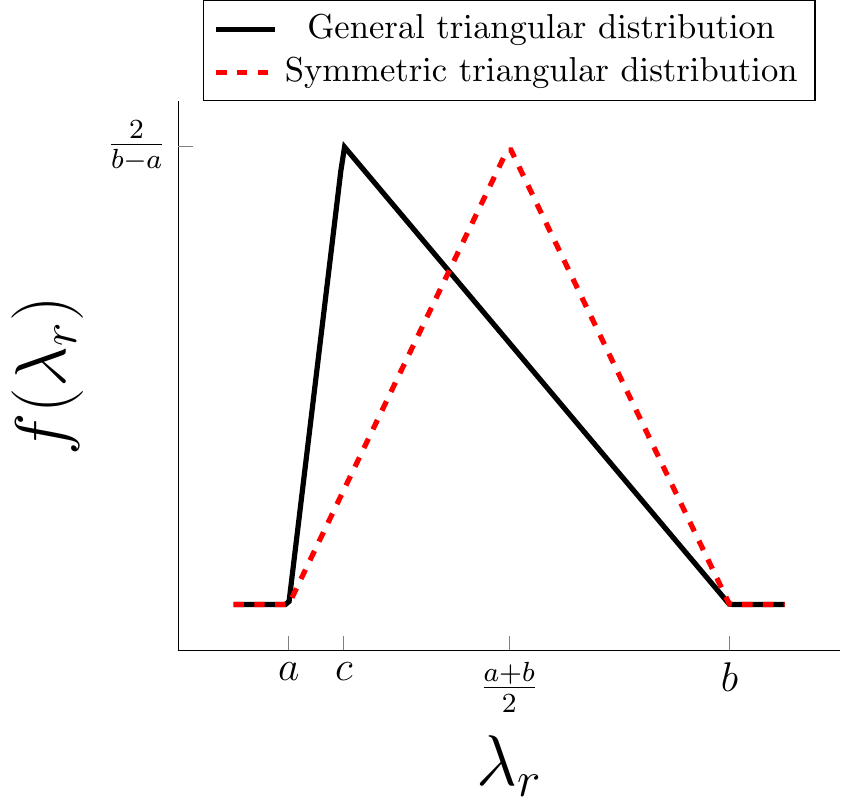}
    \subcaption[]{}
    \label{subfig:symAndGenTriDists}
  \end{subfigure}
  \caption{(a) The PDFs, $f(\lambda_r)$, for a normal distribution and two symmetric triangular distributions, with minimal value $a$ and maximal value $b$, that approximate the normal distribution. s.d. = standard deviation. (b) The PDFs, $f(\lambda_r)$, for the symmetric and general triangular distributions.}
\label{fig:distributions}
\end{figure}

Collagen fibrils in tendons are crimped when the tendon is relaxed, but straighten out as it is stretched \cite{franchi2007crimp}. We model the distribution of fibril lengths using a triangular distribution, which enables us to obtain an explicit, analytical form for the SEF. It is unlikely that such an analytical form could be derived for other PDFs; however, when it is symmetric, the triangular distribution coarsely approximates the normal distribution (see Figure \ref{subfig:normalAndSymTriDists}), and it has been shown to provide a reasonable approximation to fibril length distributions in tendons, as measured via second harmonic generation imaging \cite{bevan2018biomechanical}. An individual collagen fibril is assumed to be stress-free until becoming taut at a recruitment stretch $\lambda_r$. Once taut, it is assumed to be linearly elastic. The non-linearity of the SEF arises through the gradual recruitment of collagen fibrils \cite{kastelic1980structural}. Fibrils in the tendon are assumed to be locally coaligned. We follow the widely-used assumption that we can accurately describe soft tissue mechanics using only the isotropic invariant $I_1$ to model the tendon's non-collagenous matrix (NCM) and the anisotropic invariant $I_4$ \cite{holzapfel2001new, shearer2015new} to model the fibrils (and therefore, assume there is no dependence on $I_2$, $I_3$ or $I_5$). We decouple the contributions of the collagen fibrils and NCM in the SEF and assume that each component's contribution is proportional to its volume fraction. Finally, we assume that tendon is incompressible. Thus, the SEF, $W(I_1, I_4)$, is

\begin{equation}
    \label{eq:fullSEFForm}
    W(I_1, I_4) = (1-\phi)W_{\textrm{NCM}}(I_1) + \phi W_\textrm{coll} (I_4), 
\end{equation}
where $\phi$ is the collagen volume fraction.

To determine the form of $W_\textrm{coll}(I_4)$, we start by defining the stress exerted upon a single collagen fibril. We assume the fibrils are slack while crimped and obey Hooke's law once taut, so that the stress can be expressed as

\begin{equation}
    \label{eq:stress_on_single_fibril}
    \sigma_\textrm{fib}(\lambda, \lambda_r) =
    \begin{cases}
      0, & \lambda \leqslant \lambda_r, \\
      E \left(\frac{\lambda - \lambda_r}{\lambda_r}\right), & \lambda > \lambda_r,
    \end{cases}
\end{equation}
where $E$ is the Young's modulus of the collagen fibrils. We can determine the total (Cauchy) stress acting upon the collagen fibrils that are aligned in a given direction within a representative volume element by calculating the following integral: 

\begin{equation}
    \label{eq:fibril_stress_integral}
    \sigma_F(\lambda) = \int_0^\lambda f(\lambda_r) \sigma_\textrm{fib}(\lambda, \lambda_r) \textrm{d}\lambda_r,
\end{equation}
where $f(\lambda_r)$ represents the PDF of the recruitment stretch. We derive SEFs for two different triangular distributions: a symmetric distribution and a general distribution. We refer to them as the symmetric triangular (ST) and general triangular (GT) models, respectively. For both distributions, the first fibril becomes mechanically active at $\lambda=a$, and the last fibril becomes mechanically active at $\lambda=b$. For the ST distribution, the mode is half-way between $a$ and $b$, whereas, for the GT distribution, the mode is designated by a third parameter $c$, with $a < c < b$. The PDF for the GT distribution, $f_\textrm{gen}(\lambda_r)$, is
\begin{equation} 
    \label{eq:recruitStretch_dist}
	f_\textrm{gen}(\lambda_r) = 
	\begin{cases} 
	    0, & \textrm{ } \lambda_r<a, \\
	    \frac{2(\lambda_r - a)}{(b-a)(c-a)}, & \textrm{ } a\leqslant\lambda_r \leqslant c, \\
	    \frac{2(b-\lambda_r)}{(b-a)(b-c)}, & \text{ } c \leqslant \lambda_r \leqslant b, \\
    	0, & \text{ if } \lambda_r > b.
    \end{cases}
\end{equation}
The PDF for the ST distribution, $f_\textrm{sym}(\lambda_r)$, is obtained by setting $c = \frac{a+b}{2}$ in \eqref{eq:recruitStretch_dist} (see Figure \ref{subfig:symAndGenTriDists}).

Using \eqref{eq:stress_on_single_fibril} and the PDF of the fibril recruitment stretch distribution, we can evaluate the integral in \eqref{eq:fibril_stress_integral} analytically. Exploiting the fact that $I_4=\lambda^2$, i.e.\ $I_4$ is equal to the square of the stretch of the fibrils, we obtain
\begin{equation}
    \label{eq:stress_general_form}
    \sigma_F(I_4) = E\left(A(I_4) + B(I_4)\sqrt{I_4} + C(I_4)I_4 + \frac{D(I_4)}{2}\sqrt{I_4}\log I_4\right),    
\end{equation}
where $A(I_4),\ldots, D(I_4)$ are piecewise constants whose values depend on $I_4$ (see Appendix A). The \textit{form} of the stress, $\sigma_F$, acting on the fibrils is the same for both the ST and GT distributions; however, the piecewise constants are different in each case. In order to convert \eqref{eq:stress_general_form} into an expression for the fibrils' contribution to the SEF, we use a technique presented by Shearer \cite{shearer2015new} to write the left side of \eqref{eq:stress_general_form} in terms of $W_\textrm{coll}(I_4)$. Eventually, we obtain

\begin{IEEEeqnarray}
    {rCl}
    \label{eq:collagenSEF}
    W_{\textrm{coll}}(I_4) &=& E\bigg(\frac{A(I_4)}{2}\log I_4 + (B(I_4)-D(I_4))\sqrt{I_4} + \frac{C(I_4)}{2}I_4 + \frac{D(I_4)}{2}\sqrt{I_4}\log I_4 + G(I_4)\bigg),
\end{IEEEeqnarray}
where $G(I_4)$ is a piecewise constant that ensures the continuity of $W_\textrm{coll}(I_4)$. We further assume that the mechanical response of the NCM can be modelled by a neo-Hookean SEF \cite{shearer2015helical}, giving

\begin{IEEEeqnarray}{rCl} 
    W(I_1, I_4) = && (1-\phi)\frac{\mu}{2}(I_1-3) \nonumber \\ && + E\phi\bigg(\frac{A(I_4)}{2}\log I_4 + (B(I_4)-D(I_4))\sqrt{I_4} + \frac{C(I_4)}{2}I_4 + \frac{D(I_4)}{2}\sqrt{I_4}\log I_4 + G(I_4)\bigg),
    \label{eq:ourSEF_complete} 
\end{IEEEeqnarray}
where $\mu$ is the NCM shear modulus. A full derivation of this SEF is provided in the supplementary material. 

For an incompressible, transversely isotropic SEF that is a function of $I_1$ and $I_4$ only, the constitutive equation, in terms of the Cauchy stress, is

\begin{equation}
\label{eq:cauchyStressConstEqn}
    \bm{\sigma} = -p \mathbf{I} + 2\frac{\partial W}{\partial I_1} \mathbf{B} + 2\frac{\partial W}{\partial I_4} \mathbf{m}\otimes\mathbf{m},
\end{equation}
where $p$ is a Lagrange multiplier associated with the incompressibility constraint, and $\mathbf{m} = \mathbf{F}\mathbf{M}$ is the direction of the collagen fibrils in the deformed configuration. 

The analytical form of the SEF presented above allows the stress to be calculated rapidly compared with constitutive models that require numerical integration over the collagen recruitment stretch (e.g. \cite{aparicio2016novel,bevan2018biomechanical}). This rapidity allows the stress to be calculated millions of times within a relatively short time frame, which is exploited in our Markov Chain Monte Carlo (MCMC) approach, below.

\section{Non-Linear Optimisation}

The first two sets of stress-strain data that we fitted were experiments on mouse tail tendons collected by Goh \textit{et al}. \cite{goh2008ageing, goh2012bimodal}. We used two data sets designated as $\textrm{mtt}01\_1\_\textrm{t}5\textrm{c}$ and $\textrm{mtt}01\_1\_\textrm{t}6\textrm{b}\_\textrm{trunc}$. We shall refer to them as t5c and t6b for brevity. In these data sets, tendon specimens of length 7 mm were stretched at a displacement rate of 0.067 mm/s (a strain rate of approximately 1\%/s) \cite{goh2008ageing}. The second two sets of data were collected by Thorpe \textit{et al.} \cite{thorpe2012specialization} and were previously modelled using a different SEF by Shearer \textit{et al.} \cite{shearer2017relative}. In these, the strain rate was 5\% per second \cite{thorpe2012specialization}. We used the data sets designated as equine common digital extensor tendon (CDET) from horse number 39 and equine superficial digital flexor tendon (SDFT) from horse number 16. These data sets were selected as they have a particularly large elastic region, with the onset of failure not occurring until around 10\% strain. For brevity, we refer to them as CDET and SDFT, respectively. Each data set continues up to failure of the tendon, which is beyond the scope of our model. In order to fit only to data that is consistent with the assumptions of our model, we used the data at stretches below the point at which the maximum gradient occurs for the data collected by Goh \textit{et al}., and we used all data points up to 10\% strain for the data collected by Thorpe \textit{et al.} in accordance with \cite{shearer2017relative}.
 
\begin{figure}
    \centering
    \includegraphics{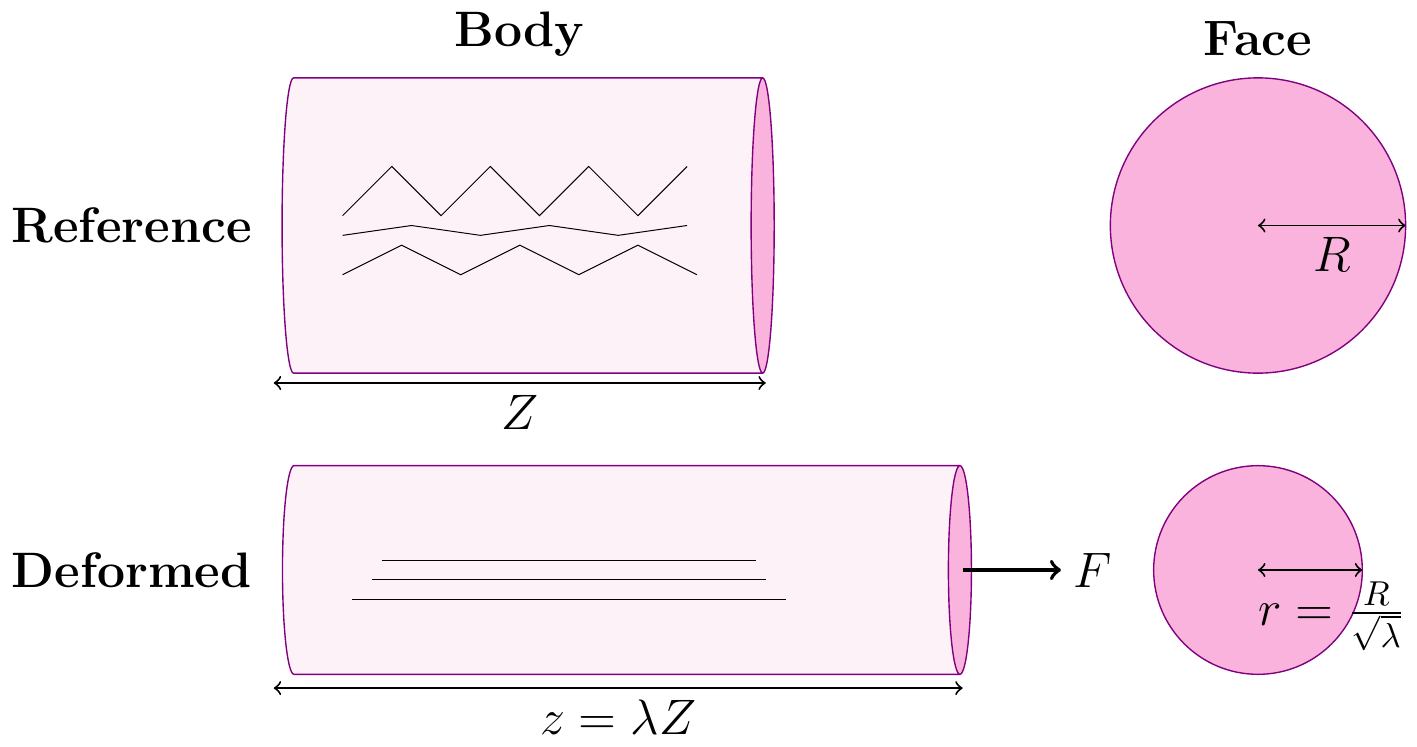}
    \caption{The tendon sample in the reference configuration (upper left) and face with normal in the $Z$-direction (upper right), and the deformed body (lower left) and face (lower right) after a force of $F \mathbf{e}_z$ is applied to the tissue.}
    \label{fig:refAndDefTendon}
\end{figure}

\begin{figure}
  \begin{subfigure}[t]{.485\textwidth}
    \centering
    \includegraphics[width=\linewidth]{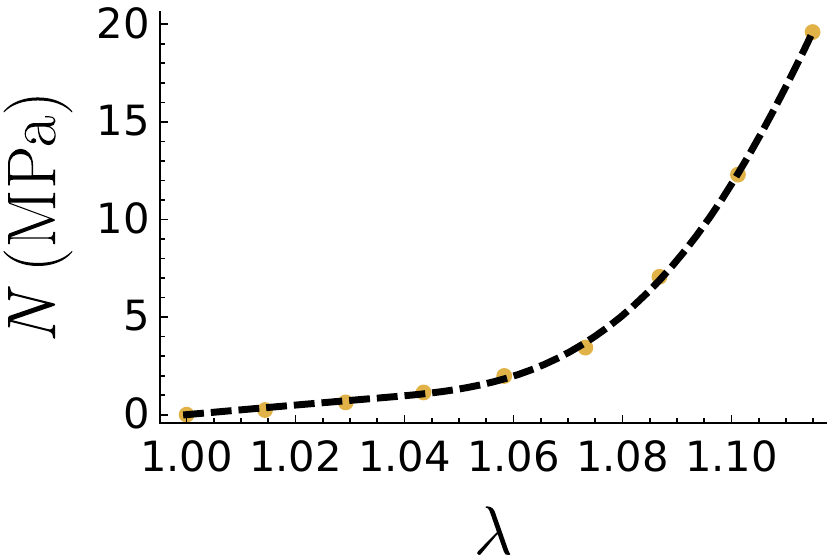}
    \caption{}
  \end{subfigure}
  \hfill
  \begin{subfigure}[t]{.485\textwidth}
    \centering
    \includegraphics[width=\linewidth]{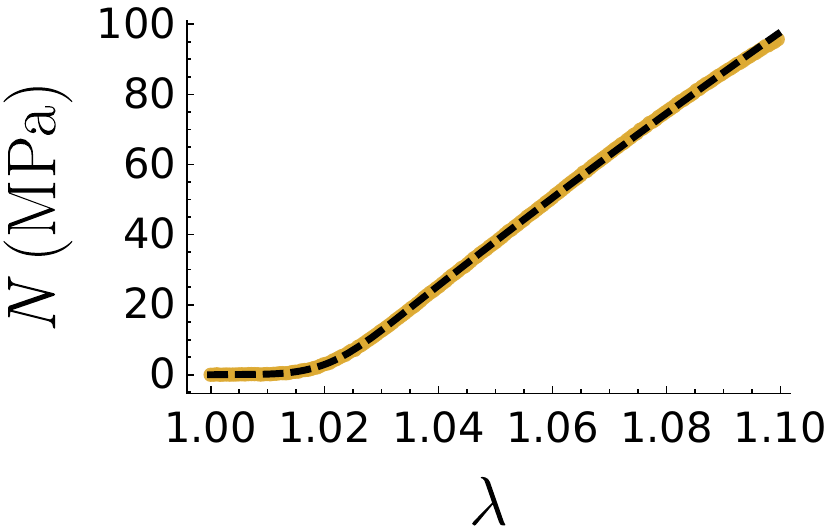}
    \caption{}
  \end{subfigure}
  \caption{Fits of (a) the ST model to the t5c data, and (b) the GT model to the CDET data. Yellow dots represent the experimental data and black, dashed lines represent model fits.}
  \label{fig:exampleMathematicaFits}
\end{figure}

To derive the constitutive equation, we assumed that a cylindrical tendon sample, described using cylindrical polar coordinates, is stretched along the axis of the aligned fibrils, which are oriented along the $Z$-axis (see Figure \ref{fig:refAndDefTendon}). For this deformation, given the assumed incompressibility and symmetry of the material, the reference and deformed coordinates, $(R, \Theta, Z)$ and $(r, \theta, z)$, are related by

\begin{equation}
    \label{eq:defAndRefCoords_Stretch}
    (r, \theta, z) = \left(\frac{R}{\sqrt{\lambda}}, \Theta, \lambda Z\right).
\end{equation}
The Cauchy stress, \eqref{eq:cauchyStressConstEqn}, gives the force acting on the deformed material per unit \textit{deformed} area; however, the quantity recorded in the experiments being modelled is the engineering stress, the force per unit \textit{reference} area, which is denoted as $N$ in Figure \ref{fig:exampleMathematicaFits}. Therefore, after substituting our SEF into \eqref{eq:cauchyStressConstEqn}, we divide the resulting expression through by $\lambda$ to obtain

\begin{equation}
    \label{eq:newModel_ES}
    N = (1-\phi)\mu\left(\lambda - \frac{1}{\lambda^2}\right) + \frac{\phi E}{\lambda}\left(A(\lambda^2) + B(\lambda^2) \lambda + C(\lambda^2) \lambda^2 + D(\lambda^2) \lambda \ln \lambda \right),
\end{equation}
where $I_4 = \lambda^2$ by \eqref{eq:defAndRefCoords_Stretch} and the assumed alignment of the fibrils.

To provide a benchmark for the fit of our model to data, we also fitted the tendon data using the commonly-used HGO model \cite{holzapfel2001new}. The HGO model was originally developed for modelling arteries and incorporates two families of collagen fibres; however, it has been adapted to study an extensive range of biological soft tissues, including tendons, and has been implemented in several finite element software packages. To model collagen in tendons, which contain one family of collagen fibres, we use the following transversely isotropic version of the HGO SEF:
\begin{equation}
    \label{eq:HGOTIversion}
    W(I_1, I_4) = \frac{c_\textrm{HGO}}{2}(I_1-3) + \frac{k_1}{2k_2}(\exp(k_2(I_4-1)^2)-1),
\end{equation}
where $c_\textrm{HGO}$ and $k_1$ are parameters with dimensions of stress and $k_2$ is a dimensionless model parameter. The engineering stress produced by this SEF \eqref{eq:HGOTIversion} is

\begin{equation}
    \label{eq:engStressHGO}
    N_\textrm{HGO} = c_\textrm{HGO} \left(\lambda - \frac{1}{\lambda^2}\right) + 2k_1 \lambda(\lambda^2-1) \exp(k_2(\lambda^2-1)).
\end{equation}
To test our model against an existing microstructural tendon model, we used the following SEF \cite{shearer2015new}:

\begin{equation}
    \label{eq:tendonModel_SEF}
    W(I_1, I_4) = (1-\phi)\frac{\mu}{2}\left(I_1-3\right) + \begin{cases}
        0, & I_4 < 1, \\
        \frac{\phi E}{6 \sin^2\theta_o} \left(4\sqrt{I_4} - 3 \ln I_4 - \frac{1}{I_4} - 3\right), & 1 \leqslant I_4 \leqslant \frac{1}{\cos^2 \theta_o}, \\
        \phi E \left(\frac{2(1-\cos^3\theta_o)}{3\sin^2\theta_o} \sqrt{I_4} - \frac{1}{2}\ln I_4 -\frac{1}{2}-\frac{\cos^2\theta_o}{\sin^2\theta_o}\log\left(\frac{1}{\cos\theta_o}\right)\right), & I_4 > \frac{1}{\cos^2\theta_o},
    \end{cases}
\end{equation}
where $\theta_o$ is the initial crimp angle of the outermost, most-crimped fibrils in the tendon's fascicles. We adapted the SEF by including a shifting parameter, $\gamma$, that corresponds to the engineering strain at which the first collagen fibril becomes mechanically active. In \eqref{eq:tendonModel_SEF}, this corresponds to replacing $\lambda$ with $\lambda - \gamma$, that is, replacing $I_4 = \lambda^2$ with $I_4 = (\lambda - \gamma)^2$. The engineering stress for this modified tendon model is

\begin{equation}
    \label{eq:engStress_tendonModel}
    N_\textrm{tendon} = (1-\phi)\mu\left(\lambda - \frac{1}{\lambda^2}\right) + \begin{cases}
        0, & \lambda < (1 + \gamma), \\
        \frac{\phi E}{3 \sin^2\theta_o} \left(2 - \frac{3}{\lambda - \gamma} - \frac{1}{(\lambda - \gamma)^3}\right), & (1 + \gamma) \leqslant \lambda \leqslant (\frac{1}{\cos \theta_o} + \gamma), \\
        \phi E \left(\frac{2(1-\cos^3\theta_o)}{3\sin^2\theta_o} - \frac{1}{\lambda - \gamma}\right), & \lambda > (\frac{1}{\cos\theta_o} + \gamma).
    \end{cases}
\end{equation}

As $\phi$, $\mu$, and $E$ only appear in the SEF \eqref{eq:ourSEF_complete} in the distinct terms $(1-\phi)\mu$ and $\phi E$, we treated $(1-\phi)\mu$ and $\phi E$ as two independent fitting parameters. Thus, the ST SEF contains four fitting parameters, $(1-\phi)\mu$, $\phi E$, $a$, and $b$. The GT SEF has an additional fitting parameter, $c$. In order to obtain physically realistic values for the parameters, we constrained them as follows: for the ST model, $0 < (1-\phi)\mu$, $0 < \phi E$, $1 < a < b$, $a < \lambda_\textrm{max}$, where $\lambda_\textrm{max}$ represents the maximum stretch in the data; for the GT model, we replaced $1 < a < b$ with $1 < a < c < b$; for the HGO model, \eqref{eq:HGOTIversion}, $0 < c_\textrm{HGO}$, $0 < k_1$, and $0 < k_2$; and for the modified tendon model, $0 \leqslant \gamma < (\lambda_\textrm{max}-1)$, $0 < \theta_o < \frac{\pi}{2}$, $(1-\phi)\mu > 0$, $\phi E > 0$.

The mean absolute error, $\Delta$, between the experimental data, $\mathbf{y}$, and simulated data, $\mathbf{\hat{y}}$, is

\begin{equation}
    \label{eq:absError}
    \Delta = \frac{1}{d}\sum_{i=1}^d |y_i - \hat{y}_i|,
\end{equation}
where $d$ is the length of the data set. Similarly, the mean relative error, $\delta$, is

\begin{equation}
    \label{eq:relError}
    \delta = \frac{1}{d}\sum_{i=1}^d \frac{|y_i - \hat{y}_i|}{|y_i|}.
\end{equation}
The values of $\Delta$ and $\delta$ when fitting each model to the four data sets are given in Table \ref{table:errors_all_fit_parameters}. To achieve the closest possible fit to the data, we ran the fitting function once and then restarted it a further four times with the estimated parameters at the end of a run used as the initial estimates for the next run. We found these restarts to have little to no effect on the quality of fit. Both versions of our SEF achieve a closer fit to the data than the microstructural tendon model for each data set. Additionally, they only perform worse than the HGO model for the relative fit to the t6b data set. Between the ST and GT models, the mean absolute and relative errors for the four data sets are similar, with the former surprisingly matching and outperforming the latter, in terms of the average relative error obtained, for the t5c and t6b data sets, respectively. This is particularly interesting because the GT model contains an additional degree of freedom. This is likely because non-linear optimisation only provides a local best fit to data and only 1000 iterations of the Nelder-Mead algorithm were performed each time the algorithm was run. Two examples of the fit of our model to the experimental data are presented in Figure \ref{fig:exampleMathematicaFits}. The supplementary material shows all sixteen fits. The parameter values found at the end of the fifth run of the non-linear fitting function are listed in Table \ref{table:mathematica_parameter_values}.
\begin{table}[ht]
	\centering

	\begin{tabular}{cccccc}
		
		Model &  & t5c & t6b & CDET & SDFT  \\ \hline
		\multirow{2}{*}{HGO} & $\delta$ & 0.242 & 0.169 & 5.50 & 1.48\\ \cline{2-6}
		& $\Delta$ (MPa) & 0.396 & 0.223 & 4.59 & 2.54\\ \hline
		\multirow{2}{*}{Tendon} & $\delta$ & 0.200 & 0.276 & 0.0967 & 0.173\\ \cline{2-6}
		& $\Delta$ (MPa) & 0.156 & 0.280 & 0.290 & 0.184\\ \hline
        ST & $\delta$ & 0.110 & 0.178 & 0.0755 & 0.149 \\ \cline{2-6}
		 & $\Delta$ (MPa) & 0.101 & 0.173 & 0.290 & 0.182\\
		\hline
		GT & $\delta$ & 0.110 & 0.181 & 0.0725 & 0.0696 \\ \cline{2-6}
		 & $\Delta$ (MPa) & 0.101 & 0.171 & 0.290 & 0.113 \\
	\end{tabular}

	\caption{Mean relative and absolute errors for the four models fitted to experimental tendon data. All values are given to three significant figures. \label{table:errors_all_fit_parameters}}
\end{table}

\begin{table}[ht]
	\centering
	\begin{tabular}{cccccc}
		
		Model &  & t5c & t6b & CDET & SDFT  \\ \hline
		\multirow{3}{*}{HGO} & $c$ (MPa) & 0 & 0 & 0 & 0\\ \cline{2-6}
		& $k_1$ (MPa) & 6.41 & 9.67 & 157 & 43.9\\ \cline{2-6}
		& $k_2$ (MPa) & 29.6 & 45.6 & 8.37 & 21.7 \\ \hline 
		\multirow{4}{*}{Tendon} & $(1-\phi)\mu$ (MPa) & 10.4 & 17.6 & 7.69 & 17.1\\ \cline{2-6}
		& $\phi E$ (MPa) & 600 & 1200 & 1330 & 778\\ \cline{2-6}
		& $\theta$ & 0.330 & 0.382 & 0.192 & 0.254\\ \cline{2-6}
		& $\gamma$ & 0.058 & 0.0461 & 0.0115 & 0.0248\\ \hline
		\multirow{4}{*}{ST} & $(1-\phi)\mu$ (MPa) & 8.36 & 14.9 & 5.31 & 16.2\\ \cline{2-6}
		& $\phi E$ (MPa) & 950 & 1760 & 1360 & 820\\ \cline{2-6}
		& $a$ & 1.04 & 1.03 & 1.01 & 1.02\\ \cline{2-6}
		& $b$ & 1.16 & 1.15 & 1.03 & 1.07\\ \hline \multirow{5}{*}{GT} & $(1-\phi)\mu$ (MPa) & 8.36 & 15.1 & 4.78 & 12.3\\ \cline{2-6}
		& $\phi E$ (MPa) & 931 & 782 & 1360 & 828\\ \cline{2-6}
		& $a$ & 1.04 & 1.03 & 1.01 & 1.01\\ \cline{2-6}
		& $c_{\textrm{mode}}$ & 1.10 & 1.09 & 1.02 & 1.06\\ \cline{2-6}
		& $b$ & 1.16 & 1.09 & 1.03 & 1.06\\
	\end{tabular}
	\caption{Parameter values (to three significant figures) for each model's best fit to the four data sets. To avoid confusion with $c$ in the HGO model, the modal fibril length in the GT model is written as $c_{\textrm{mode}}$.} \label{table:mathematica_parameter_values}
\end{table}

\section{Markov chain Monte Carlo}

Through non-linear optimisation, we have found the best fit to experimental data local to the algorithm's initial guesses for the parameter values; however, this approach does not quantify the uncertainty in the parameter values. Uncertainties arise for a number of reasons, including observational noise in the experimental stress-strain data. To address this, we apply a Bayesian framework to the same problem studied with the optimisation approach and estimate the likely ranges of the true values of the model's parameters. Using the posterior distributions we obtain from the algorithm, we can estimate likely parameter values and quantify the uncertainty in those estimates.
    
The goal of Bayesian statistics is, given new data, to update any prior knowledge about the values of a model's parameters via the likelihood of a particular parameter vector $\bm{\theta}$ (the vector of constitutive and structural parameters in our SEF) producing the observed (experimental) data $\mathbf{y}$. Through this, we obtain what is known as the posterior probability distribution of $\bm{\theta}$, $\pi(\bm{\theta}|\mathbf{y})$, which is related to $\pi_0(\bm{\theta})$, the prior probability of $\bm{\theta}$, and the likelihood via Bayes' rule:
	
\begin{equation}
    \label{eq:BR_propVersion}
	\pi(\bm{\theta}|\mathbf{y}) \propto L(\mathbf{y}|\bm{\theta})\pi_0(\bm{\theta}),
\end{equation}
where $L(\mathbf{y}|\bm{\theta})$ denotes a function that is proportional to the likelihood density. The posterior is only known up to a constant of proportionality, which often cannot be explicitly computed. Under those circumstances, a common method to characterise the posterior distribution is to sample from it using numerical methods such as MCMC. 
	
Monte Carlo methods can be used to estimate expectations with respect to a particular measure, for example $\pi(\bm{\theta}|\mathbf{y})$; however, for Bayesian inverse problems, we cannot usually directly sample from the posterior distribution. Instead, we can indirectly sample from the posterior using MCMC methods, which construct an ergodic Markov chain whose unique stationary density is equal to the posterior. Monte Carlo estimates taken with respect to this Markov chain can be shown to converge to expectations taken with respect to the posterior distribution. Initially, Markov chains do not sample from the stationary distribution and values proposed in the MCMC algorithm are dependent on the chains' starting position. This initial period is called the burn-in phase, the size of which depends on the quality of the initial guess, and the rate of mixing of the Markov chain. We do not include samples from the burn-in phase when calculating MCMC estimates, or when visualising the posterior distribution.

\subsection{Hierarchical Bayesian approach and conjugate priors}

In Section \ref{sec:themodel}, we derived a deterministic SEF. Now, in order to derive the likelihood function for this modelling problem, we assume that the observed data is given by the model output perturbed by some noise. We choose the standard modelling assumption that the noise is additive, mean-zero, Gaussian and independently and identically distributed (IID), giving us a diagonal covariance matrix for which the entries are equal to the observational noise variance $\sigma^2$. This gives us the following statistical model for our observations:

\begin{equation}
    \label{eq:DataEquationNoise}
    \mathbf{y} = \mathbf{M}(\bm{\theta}) + \bm{\eta}, \hspace{1.5cm} \bm{\eta}\sim \mathcal{N}(\mathbf{0},\sigma^2\mathbf{I}_{d}),
\end{equation}
where $d$ is the length of $\mathbf{y}$, $\mathbf{I}_{d}$ is a $d\times d$ identity matrix, $\mathcal{N}(\mathbf{0}, \sigma^2 \mathbf{I}_d)$ represents a normal distribution with mean $\mathbf{0}$ and covariance matrix $\sigma^2 \mathbf{I}_d$, $\mathbf{y} \in \mathbb{R}^d$, and $\mathbf{M}(\bm{\theta})$ denotes the output of the model given input values $\bm{\theta}$. From \eqref{eq:DataEquationNoise}, we derive the likelihood, which is induced by the statistical model on the left of this equation, once the noise is assumed to be a zero-mean, homoscedastic, multivariate, Gaussian random variable:
\begin{equation}
\label{eq:likelihood}
L(\mathbf{y}|\bm{\theta},\sigma^2) = \frac{1}{(\sigma \sqrt{2\pi})^d} \exp \left ( -\frac{1}{2\sigma^2}\|\mathbf{y} - \mathbf{M}(\bm{\theta})\|^2_2 \right ).
\end{equation}
This is sufficient if we have a clear idea of the value of the observational noise variance $\sigma^2$, but in practice this is rarely the case. The value of $\sigma^2$ can be very important, potentially causing under- or over-fitting. Therefore we take a hierarchical Bayesian approach and assign a prior distribution to $\sigma^2$. \textit{A priori}, we assume that the parameters are independent of one another, so the joint prior distribution is  the product of the parameters' individual prior distributions. That is,

\begin{equation}
    \label{eq:Priors_Noise}
    \pi_0(\bm{\theta},\sigma^2) = \pi_0({\bm \theta}) \pi_0(\sigma^2) = \pi_0(\theta_1)\cdots\pi_0(\theta_h)\pi_0(\sigma^2), 
\end{equation}
where $h$ denotes the length of $\bm{\theta}$. By \eqref{eq:BR_propVersion} and \eqref{eq:Priors_Noise},

\begin{equation}
    \label{eq:posteriorNoise}
    \pi(\bm{\theta},\sigma^2|\mathbf{y}) \propto L(\mathbf{y}|\bm{\theta},\sigma^2)\pi_0(\theta_1)\cdots\pi_0(\theta_h)\pi_0(\sigma^2).
\end{equation}

Using a conjugate prior for $\sigma^2$, we avoid having to infer $\sigma^2$ explicitly by integrating out the dependence of the posterior distribution with respect to $\sigma^2$ since this integral can be computed analytically. In this instance, our likelihood function is a Gaussian PDF, so an appropriate conjugate prior for the observational noise variance is an inverse-gamma distribution. After multiplying the likelihood function by the product of the prior densities \eqref{eq:Priors_Noise}, we arrive at the posterior distribution. The marginal distribution on the model parameters can then be derived by integrating out $\sigma^2$, giving a Student's t-distribution multiplied by the prior density on the remaining unknowns:

\begin{IEEEeqnarray}{rCl}
\label{eq:posteriorAfterConjugatePrior}
    \pi(\bm{\theta}|\mathbf{y})
    &\propto& \textrm{t}_{2\alpha_\sigma}\left(\mathbf{y};\mathbf{M}(\bm{\theta}),\frac{\beta_\sigma}{\alpha_\sigma}\mathbf{I}_d\right)\pi_0(\bm{\theta}),
\end{IEEEeqnarray}
where we define $t_{2\alpha_\sigma}(\mathbf{*};\bm{\gamma}, \bm{\psi})$ as the posterior predictive density of $\mathbf{*}$ according to a Student's t-distribution of $2\alpha_\sigma$ degrees of freedom with mean $\bm{\gamma}$ and covariance matrix $\bm{\psi}$, and where $\alpha_\sigma, \beta_\sigma >0$ are parameters of the hyperprior on $\sigma^2$. For the implementations of the RWM algorithm detailed below, we set the values of the hyperparameters to be $\alpha_{\sigma} = 3$ and $\beta_{\sigma} = 0.3$. A full derivation of the posterior predictive is provided in the supplementary material.

\subsection{Random walk Metropolis algorithm}

We cannot integrate \eqref{eq:posteriorAfterConjugatePrior} analytically and determine the normalisation constant; therefore, we choose to characterise the posterior predictive by sampling from the target distribution using the random walk Metropolis (RWM) algorithm. This method enables us to construct an ergodic Markov chain with invariant density equal to $\pi(\bm{\theta}| \mathbf{y})$. We can then use the computed Markov chain for Monte Carlo estimates and to visualise the target distribution. Producing a Markov chain of length $n$ and using $\mathcal{U}(0, 1)$ to denote a uniform distribution between zero and one, the RWM algorithm is given in Algorithm \ref{algo:RWM}.

\begin{algorithm}[H]
\SetAlgoLined
\KwResult{Estimate of the posterior distribution, $\pi(\bm{\theta})$}
 \textbf{Input}:\
 $\pi_0(\theta_1)$, \ldots, $\pi_0(\theta_h)$\;
 Starting parameter vector, $\bm{\theta}_0$\;
 \For{$i = 1$, \ldots, $n$}{
  Propose $\bm{\theta}^* \sim \mathcal{N}(\bm{\theta}_{i-1}, \bm{\Sigma})$\;
  Calculate $\kappa = \textrm{min}\left(1, \frac{\textrm{t}_{2\alpha_\sigma}\left(\mathbf{y};\mathbf{M}(\bm{\theta}^*),\frac{\beta_\sigma}{\alpha_\sigma}\mathbf{I}_d\right)\pi_0(\bm{\theta}^*)} {\textrm{t}_{2\alpha_\sigma}\left(\mathbf{y};\mathbf{M}(\bm{\theta}_{i-1}),\frac{\beta_\sigma}{\alpha_\sigma}\mathbf{I}_d\right)\pi_0(\bm{\theta}_{i-1})}\right)$\;
  Generate $u \sim \mathcal{U}(0, 1)$\;
  \eIf{$u \leqslant \kappa$}{
   $\bm{\theta}_{i} = \bm{\theta}^*$\;
   }{
   $\bm{\theta}_{i} = \bm{\theta}_{i-1}$\;
  }
  Set $i = i+1$\;
 }
\caption{RWM \label{algo:RWM}}
\end{algorithm}

The covariance matrix of the proposal distribution, $\bm{\Sigma}$, affects the efficiency of RWM algorithms. It controls the scale of the proposal variance and the correlations between coordinates in the sampled vector. Adaptive random walk algorithms allow us to adapt $\bm{\Sigma}$ to optimise efficiency. To address the scale and correlation of the sample vector, we adapt the covariance matrix to $\bm{\Sigma} = \beta^2 \bm{\zeta}$, where $\beta^2$ is a scaling parameter, with $\beta>0$, and $\bm{\zeta} \in \mathbb{R}^{h\times h}$ is the covariance matrix of the parameters constructed from a chosen set of parameter vectors. 

Regarding scale, it has been shown that the optimal acceptance rate for multivariate RWM is 0.234 \cite{roberts2001optimal}. For a given value of $\bm{\zeta}$, $\beta$ can be tuned to achieve an acceptance rate close to this value. Small values of $\beta$ lead to a proposal density closely concentrated around the current state, which leads to a high acceptance rate but slow exploration. Conversely, large values of $\beta$ lead to a diffuse proposal density where sampled vectors are likely to be in the tails of the posterior distribution, leading to low acceptance rates and therefore slow exploration.

Efficient proposal distributions reflect the correlation structures in the target density. For instance, if the probability density is concentrated close to a lower dimensional manifold, then proposal distributions which favour bigger moves in the directions parallel to the manifold will lead to faster exploration than isotropic proposal distributions. We do not know the correlation structure of the target \emph{a priori}, but this can be learned through initial exploration with an isotropic proposal distribution.

We employed an adaptive RWM method, recalculating $\bm{\Sigma}$ after every block of 500 samples. To construct $\bm{\zeta}$, we used the position of the Markov chains over the last 10,000 samples in the chain. To ensure $\bm{\Sigma}$ was positive definite during the algorithm, we regularised by adding the identity matrix multiplied by a small number, \num{1e-5}, to $\bm{\zeta}$ whenever it was recalculated. We let the value of $\beta^2$ depend on the acceptance rate within a block, $\alpha_\textrm{block}$. The conditions for updating $\beta^2$ at the end of each block were

\begin{itemize}
    \item $\alpha_{\textrm{block}} < \alpha_{\textrm{LowerTol}}$: multiply $\beta^2$ by $0.95^2$;
    \item $\alpha_{\textrm{LowerTol}} \leqslant \alpha_{\textrm{block}} \leqslant  \alpha_{\textrm{UpperTol}}$: keep $\beta^2$ at the same value;
    \item $\alpha_{\textrm{UpperTol}} < \alpha_{\textrm{block}}$: multiply $\beta^2$ by $1.05^2$,
\end{itemize}
where $\alpha_\textrm{LowerTol}$ and $\alpha_\textrm{UpperTol}$ denote the lower and upper bounds of the allowed acceptance rates for the algorithm, which we set equal to 0.184 and 0.284, respectively $(0.234\pm0.05)$. The tolerances account for the range of acceptance rates for which an RWM algorithm is assumed to run efficiently enough. We must stop iterating $\beta^2$ and $\bm{\zeta}$ at some point in the algorithm, since adaptive MCMC algorithms must satisfy the property of diminishing adaptation in order to maintain ergodicity \cite{roberts2007coupling}. We stopped adaptation of $\bm{\Sigma}$ at the end of the burn-in phase, which consisted of the first 500,000 samples of the Markov chain. Convergence was checked through the repeatability and smoothness of the computed histograms, but more formal methods to quantify convergence are available~\cite{gelman1992inference}.

\section{Application of Bayesian methods to tendon deformation}

Before running the RWM algorithm, we transformed the parameters so their support extended over the whole of $\mathbb{R}$ because sampling parameters whose support matches that of the proposal distributions improves efficiency. Here we discuss the approach for the ST model. The GT model is discussed in the supplementary material. Each element of the parameter vector, $\bm{\psi} = [(1-\phi)\mu, \phi E, a, b]$, is non-negative, and the uncertain parameters $a, b$ must satisfy $a>1$ and $a<b$. These two conditions give rise to the natural choice of parameters for inference $a-1 > 0$ and $b-a > 0$. Along with the other non-negative, uncertain parameters, we assigned log-normal priors to ensure well-posedness. Taking the logarithm of these parameters, we obtained the parameter vector $\bm{\theta} \in \mathbb{R}^4$, where
\begin{equation}
    \label{eq:transfParams}
    \bm{\theta} = \left(\begin{array}{c}
         \nu  \\
         \eta \\
         \tau \\
         \rho
    \end{array}\right) =
    \left(\begin{array}{c}
         \log((1-\phi)\mu)  \\
         \log(\phi E) \\
         \log(a-1) \\
         \log(b-a)
    \end{array}\right) = 
    \textrm{T}(\bm{\psi}),
\end{equation}
where $\textrm{T}(\bm{\psi})$ represents an invertible, non-linear transformation of the target parameters $\bm{\psi}$. This transformation, in turn, leads to a transformation of the likelihood and posterior distributions. When performing RWM on the transformed parameters, $\bm{\theta}$, the target density is given by the pullback $\tilde{\pi}$ of the posterior $\pi(\bm{\psi}|\bm{y})$ through the map $T$, which has density
\begin{equation}
    \tilde{\pi}(\bm{\theta}) = \pi(T^{-1}(\bm{\theta})|\bm{y}).|\det D_{T^{-1}}(\bm{\theta})|,
\end{equation}
where $D_{T^{-1}}(\bm{\theta})$ is the Jacobian of $T^{-1}$. The value of this additional factor is detailed in the supplementary material. As we assigned a log-normal prior to their exponents, each parameter in $\bm{\theta}$ has a normal prior distribution. The supplementary material details how the two parameters of the log-normal prior, and, thus, the mean and variance of the corresponding normal prior, were chosen for $(1-\phi)\mu$, $\phi E$, $a - 1$, and $b - a$. 

\begin{figure}[htbp]

\includegraphics[trim = 20 59 40 5, clip]{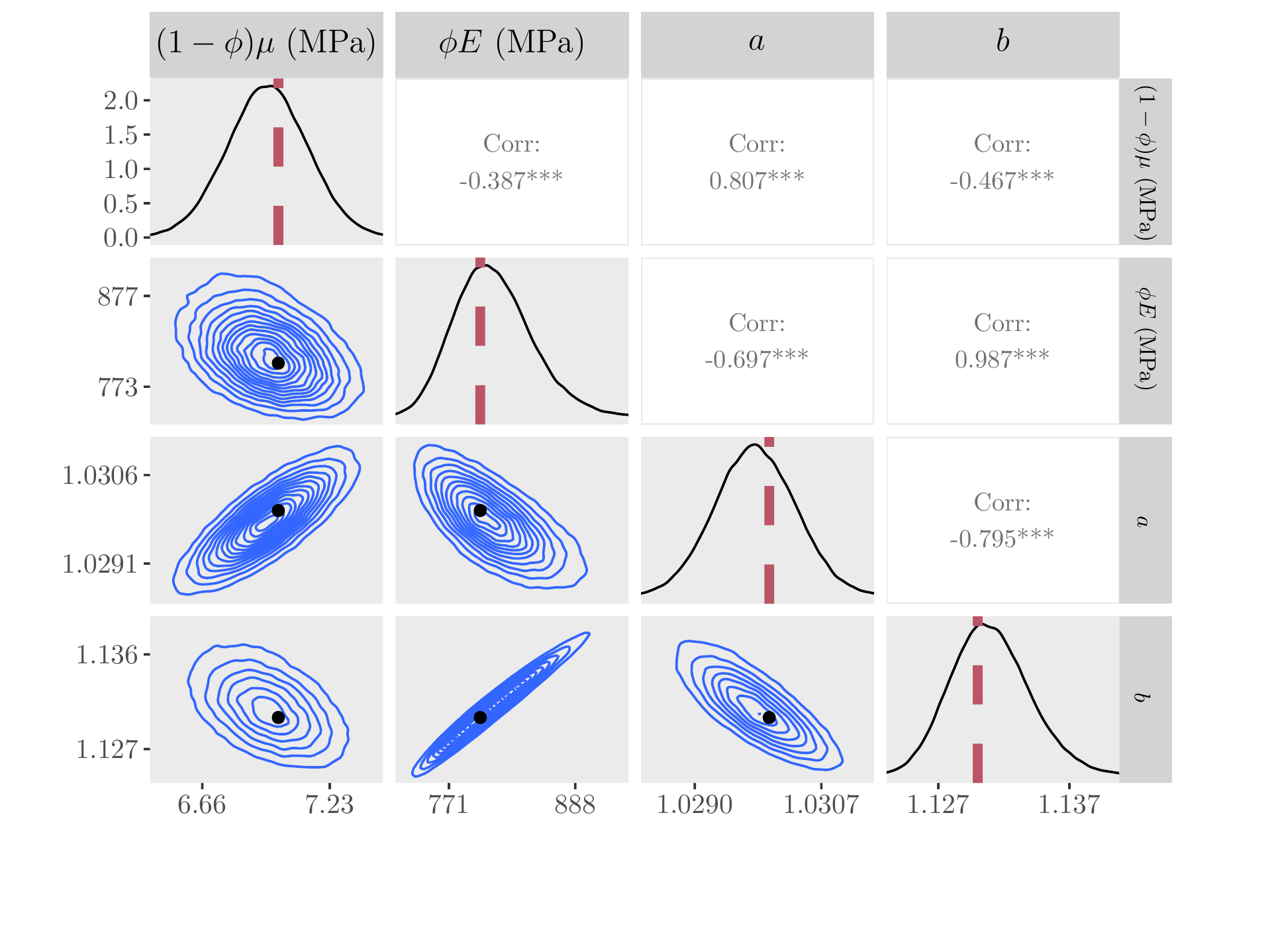}
\caption{Plots of the posterior distributions calculated using the RWM algorithm on synthetic data. Main diagonal: marginal posteriors. Lower half: two-dimensional contour plots of the joint distributions. Upper half: posterior correlations between parameters. The parameter values used to create the synthetic data are represented by a red line on the posteriors and a black dot on the contour plots. For the correlation values, three asterisks represent $p < 0.001$. In order to create this figure, the 1 million samples were thinned by a factor of ten.}
\label{fig:synth_correlogram}
\end{figure}

\begin{figure}[htbp]
\centering
\includegraphics[trim = 25 61 45 23, clip]{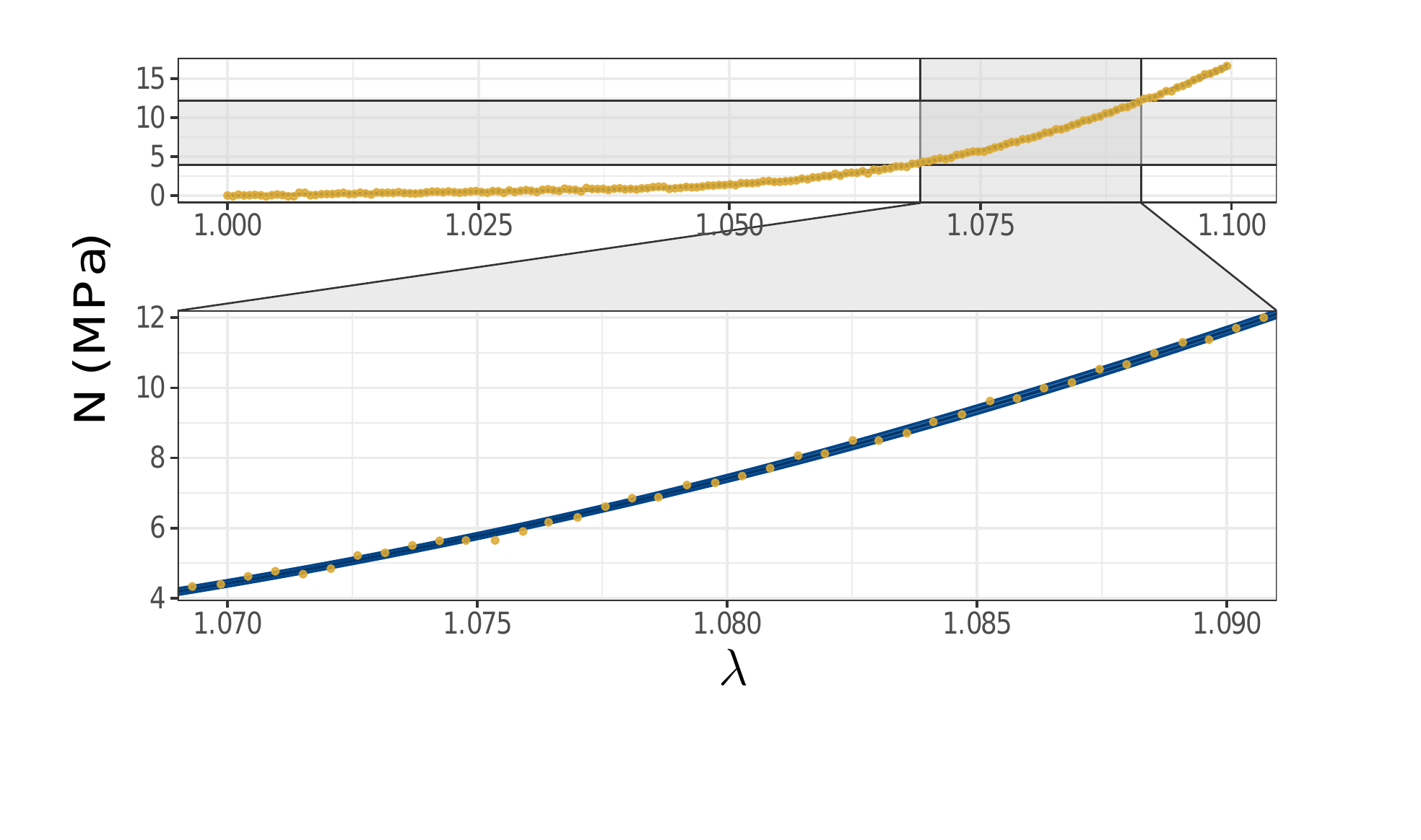}
\caption{The $5\sigma$ confidence band (blue) around the mean of the predicted stresses from 50,000 parameter vectors from the Markov chains against the synthetic data (yellow dots).}
\label{fig:synth_meanAndCB}
\end{figure}

To validate our approach and its implementation, we first used it on a synthetic data set created with a chosen parameter vector for the ST model, before moving on to the CDET and SDFT data, for which we used both the ST and GT models. We also ran the ST algorithm on the data collected by Goh \textit{et al}, as discussed in the supplementary material. All runs of the algorithm described below included a burn-in phase of 500,000 samples that were not included in the results.

\subsection{Synthetic data}
Fitting to synthetic data acts as a proof-of-concept that enables us to study the posterior distributions when we know the `true' parameter values associated with the data. To create the synthetic stress values, we inputted the same set of strains as the SDFT data set and the parameters $[(1-\phi)\mu, \phi E, a, b] = [7 \textrm{ MPa}, 800 \textrm{ MPa}, 1.03, 1.13]$ into the SEF. To simulate experimentally collected data, we added IID mean zero Gaussian noise to the stresses, and to test the algorithm rigorously, we made the synthetic data noisier than the real data by choosing a variance of 0.01 for the noise. We produced a Markov chain of 1.5 million samples for the synthetic data.

The marginal posterior distributions and the two-dimensional joint distributions of the parameters that were obtained when fitting the synthetic data are shown in Figure \ref{fig:synth_correlogram}. Although they do not align exactly with the modes of their respective posteriors, the parameter values used to create the synthetic data are not located in the tails of the posterior, but lie in regions of relatively high posterior probability. The smoothness of the empirical distribution also implies that the algorithm sampled efficiently.

Figure \ref{fig:synth_meanAndCB} shows the fit to the data and a $5\sigma$ confidence band around the mean predicted stresses for a sample of 50,000 parameter vectors. A close fit is achieved for the whole stress-strain curve. This synthetic experiment demonstrates that accurate estimates of the constitutive parameters, and the uncertainty in those estimates, can be derived through a Bayesian framework, and characterised using our tuned adaptive RWM algorithm.

\subsection{SDFT and CDET data: ST model}

\begin{figure}
\centering
\includegraphics[trim = 25 59 40 5, clip]{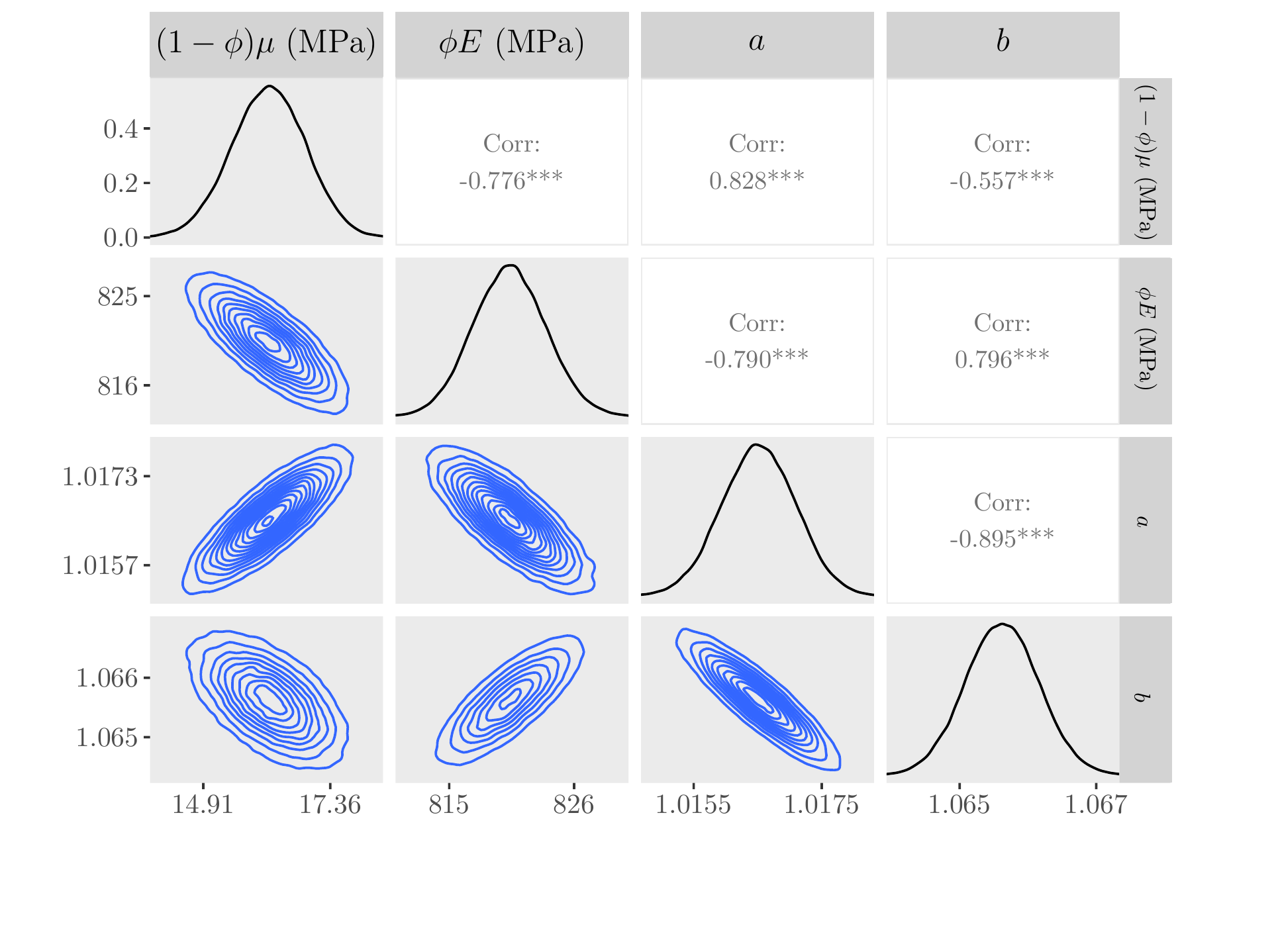}
\caption{Approximate posteriors and contour plots of the parameters for the SDFT data. Samples were thinned by a factor of ten.}
\label{fig:SDFT_correlogram}
\end{figure}

\begin{figure}
\centering
\includegraphics[trim = 25 61 45 23, clip]{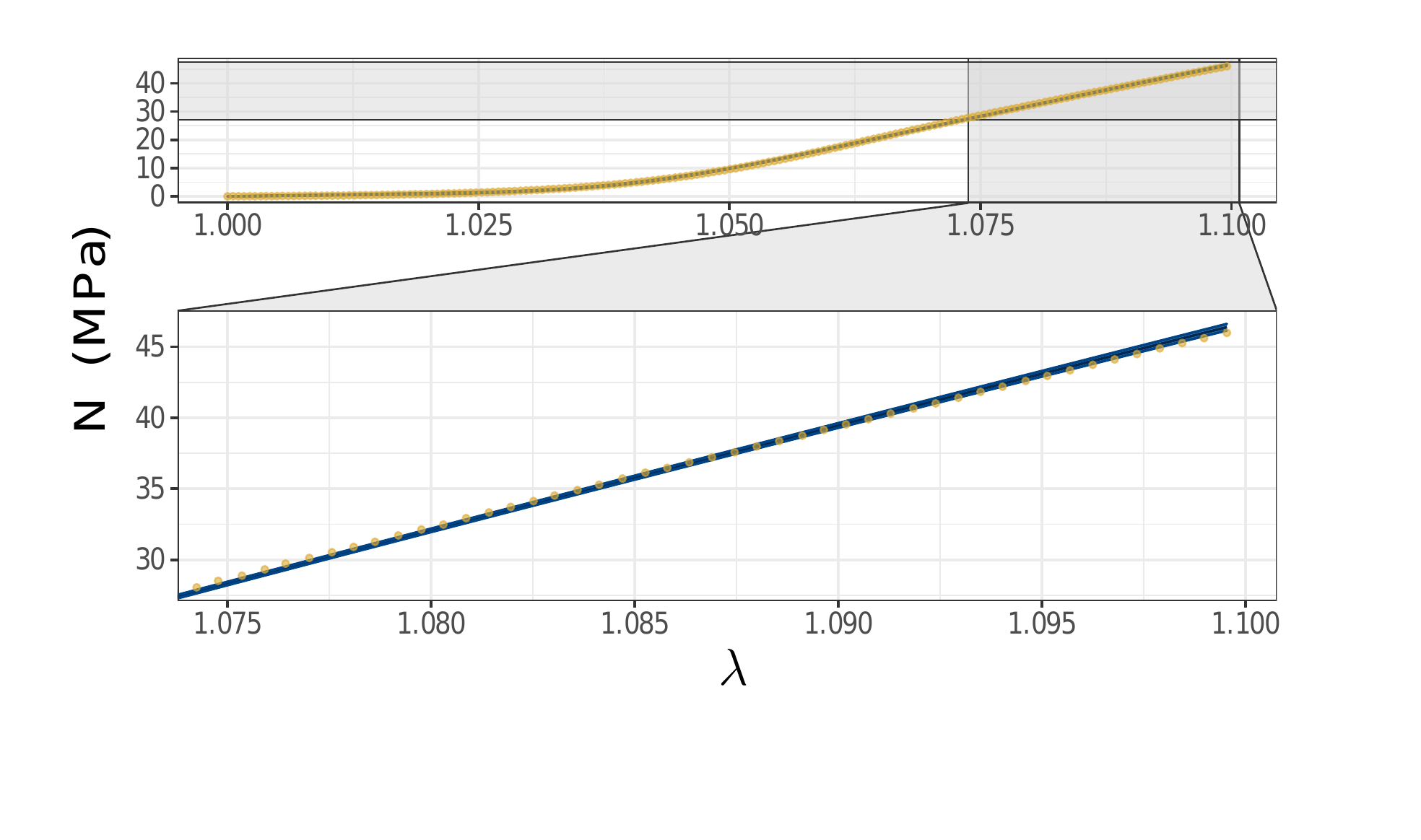}
\caption{The $5\sigma$ confidence band (blue) around the mean of the predicted stresses from 50,000 parameter vectors from the Markov chains against the SDFT data (yellow dots).}
\label{fig:SDFT_meanAndCB}
\end{figure}

We now analyse the algorithm's predictions when we fit the ST model to the high-resolution tendon data collected by Thorpe \textit{et al}. We used the ST model based on the assumption that the fibril length distribution is symmetric; however, we relax this assumption in Section \ref{sec:GT}. We produced a Markov chain of 1.5 million samples for the SDFT data and 10.5 million for the CDET data. The additional nine million samples for the CDET data were performed to help produce smoother posteriors as their irregular shapes caused slower mixing of the Markov chains. For the SDFT data, Figure \ref{fig:SDFT_correlogram} contains the estimated posteriors and contour plots obtained from the RWM algorithm and Figure \ref{fig:SDFT_meanAndCB} shows a confidence band of $5\sigma$ around the mean stress-strain curve of 50,000 parameter vectors plotted against the data. As in the synthetic example, the empirical posterior distribution is smooth, implying good convergence. For the structural parameter $\phi E$, we have a physically realistic posterior: the 95\% credible interval for $\phi E$ is 814-827 MPa, which is feasible compared to literature values for $\phi$ and $E$ in tendon \cite{svensson2012mechanical}. The stretches at which the first and last collagen fibrils tauten are also realistic. There are parameters with strong positive correlations: $(1-\phi)\mu$ and $a$, and $\phi E$ and $b$. These indicate that to replicate the experimental data closely, the NCM must be stiffer if collagen fibrils are slack for longer, and the fibrils must be stiffer if fewer are mechanically active. Likewise, strong negative correlations between $a$ and $b$ and $(1-\phi)\mu$ and $\phi E$ indicate that the final fibril must be taut sooner if the first is slack for longer, and the fibrils must be stiffer if the NCM is more compliant. These are all physically reasonable correlations, demonstrating the benefit of full posterior characterisation as opposed to traditional optimisation.

Figure \ref{fig:SDFT_meanAndCB} demonstrates that the algorithm identifies parameter vectors that fit the SDFT data closely. For all stretches, the data lies close to the mean stress, and either within or close to the $5\sigma$ confidence band that is narrower than for the noisier synthetic data. Again, the confidence band is consistently sized, demonstrating the model's ability to quantify how different microstructural components and phenomena (the NCM and the gradual tautening of collagen fibrils) influence the macroscopic mechanical response of the tendon. As the strain nears 10\%, however, the model's predicted stresses are slightly higher than the experimental data, indicating a degree of discrepancy between the model and data. This could be due to damage to some fibrils as the stretch nears 10\% strain, contradicting an assumption of the model. To achieve the best fit to the data overall, while retaining the linearity of the model in region III of the stress-strain curve, some underestimates of the experimental stress occur at smaller stretches to compensate for the overestimates as the strain approaches 10\%.

\begin{figure}
\includegraphics[trim = 25 59 40 5, clip, width=0.98\textwidth]{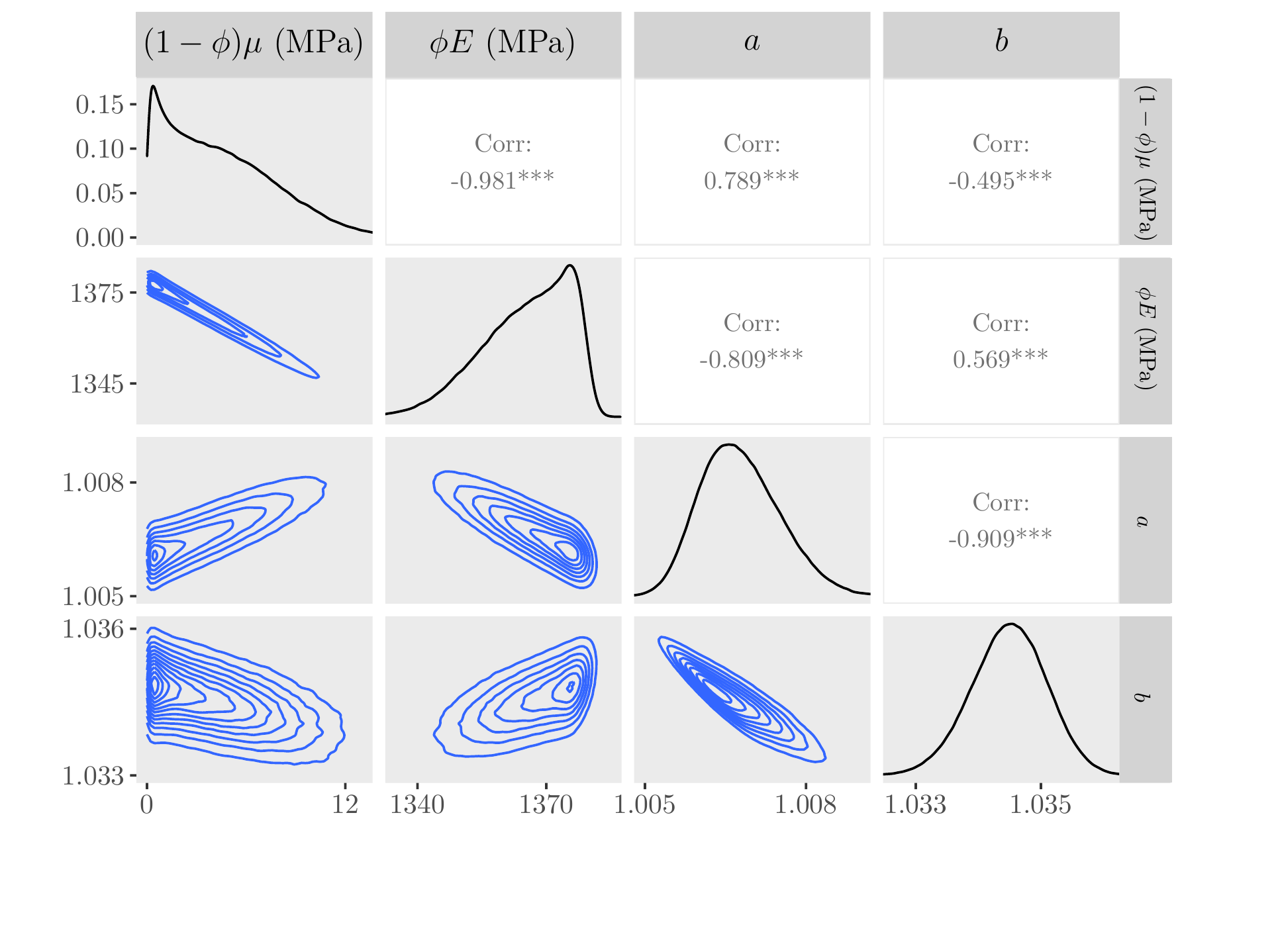}
\caption{Approximate posteriors and contour plots of the parameters for the CDET data. Samples were thinned by a factor of ten.}
\label{fig:CDET_correlogram}
\end{figure}

\begin{figure}
\centering
\includegraphics[trim = 25 61 45 19, clip, width=0.98\textwidth]{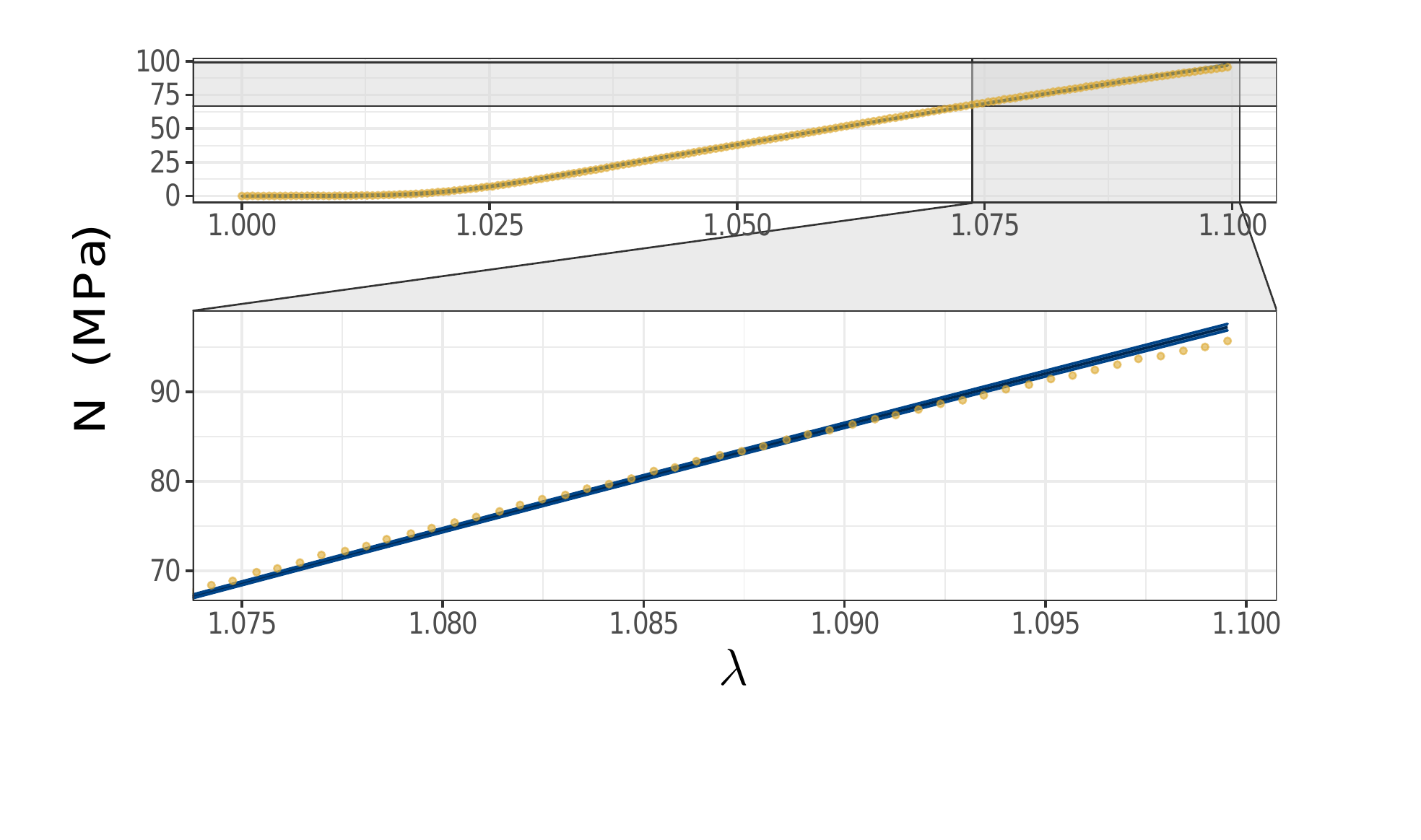}
\caption{The $5\sigma$ confidence band (blue) around the mean of the predicted stresses from 50,000 parameter vectors from the Markov chains against the CDET data (yellow dots).}
\label{fig:CDET_meanAndCB}
\end{figure}

For the CDET data, the parameter $(1-\phi)\mu$ possesses a high predicted posterior probability mass close to zero (see Figure \ref{fig:CDET_correlogram}), with a long tail for larger values. Due to the strong negative correlation between $(1-\phi)\mu$ and $\phi E$, the shape of the marginal distribution for $\phi E$ is also affected. The shape of the posterior for $(1-\phi)\mu$ likely occurs because few data points lie in the toe region, with the proposed values of $a$ being close to one, meaning that the collagen fibrils dominate the response to the deformation even at small stretches. As the density lies close to zero for one of the parameters, taking the logarithm results in a curved posterior, whose global covariance structure is less informative for making effective proposals. Therefore, in order to obtain smoother posteriors, 10 million samples were taken. Alternative approaches would be to use more sophisticated methods such as the Metropolis-Adjusted Langevin algorithm or Hamiltonian Monte Carlo \cite{girolami2011Riemann}. The posteriors for $a$ and $b$ are smooth, implying a good level of convergence to the posterior distributions. Furthermore, close fits to the data from the sampled parameters are still achieved (see Figure \ref{fig:CDET_meanAndCB}). With a 95\% credible interval of 1340-1380 MPa, we again obtain physically reasonable estimates of $\phi E$ \cite{svensson2012mechanical}.

\subsection{SDFT and CDET data: GT model}\label{sec:GT}
\begin{figure}
\includegraphics[trim = 25 59 40 5, clip, width=0.98\textwidth]{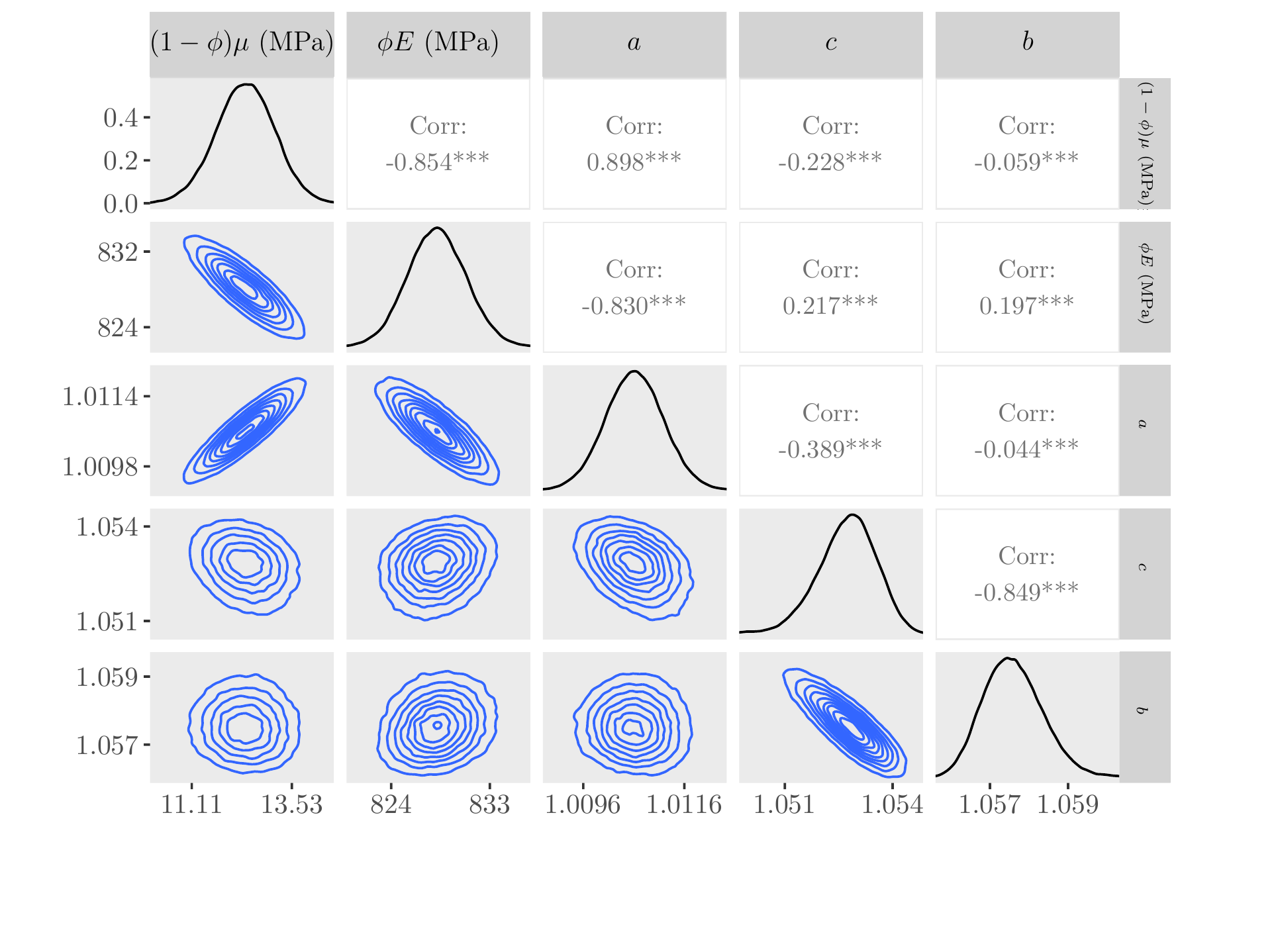}
\caption{Approximate posteriors and contour plots of the parameters of the GT model for the SDFT data. Samples were thinned by a factor of ten.}
\label{fig:SDFT_GT_correlogram}
\end{figure}

\begin{figure}
\includegraphics[trim = 25 59 40 5, clip, width=0.98\textwidth]{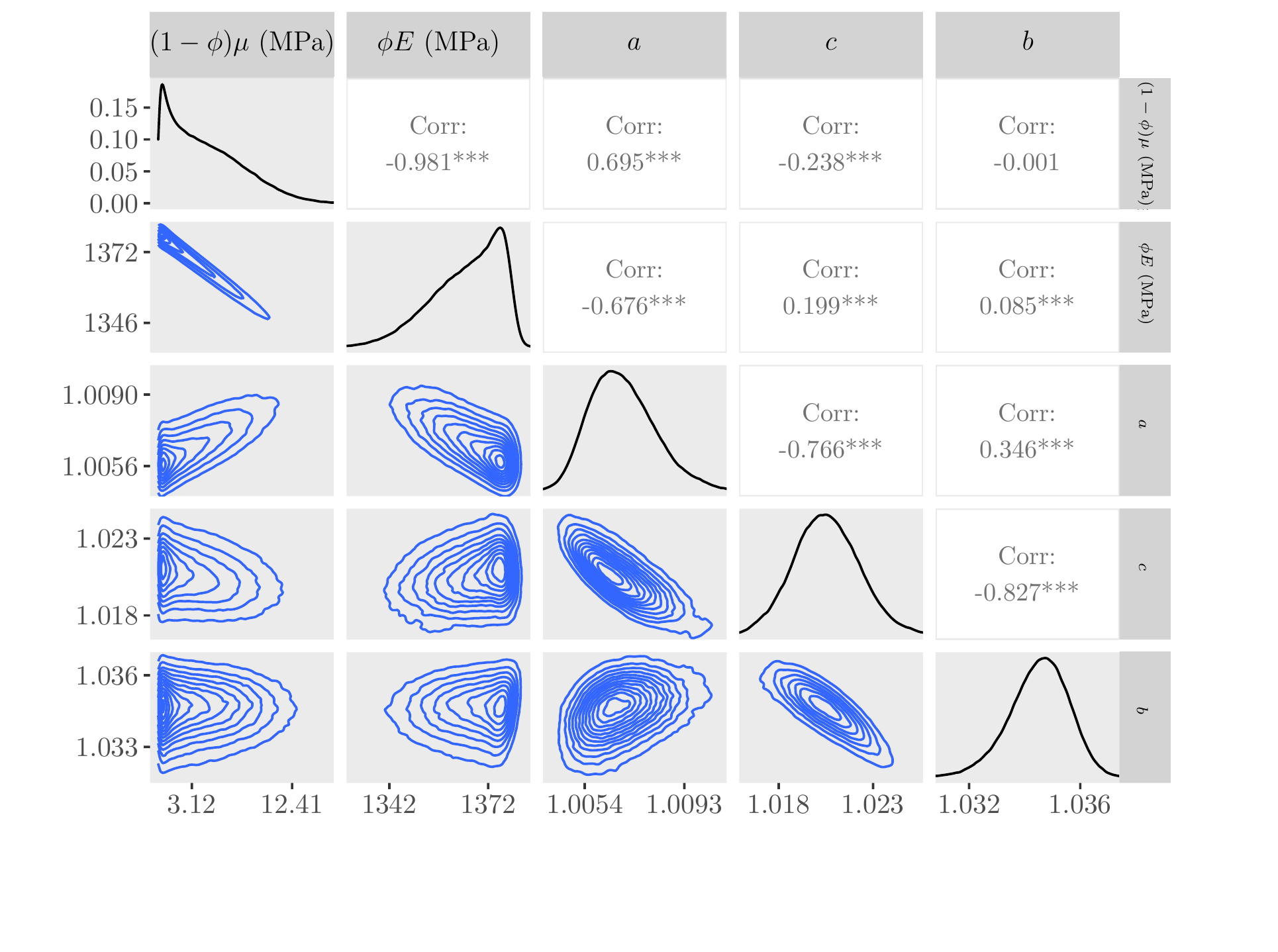}
\caption{Approximate posteriors and contour plots of the parameters of the GT model for the CDET data. Samples were thinned by a factor of ten.}
\label{fig:CDET_GT_correlogram}
\end{figure}

\begin{figure}
   \begin{subfigure}[t]{.485\textwidth}
    \centering
    \includegraphics[trim = 17 47 57 19, clip]{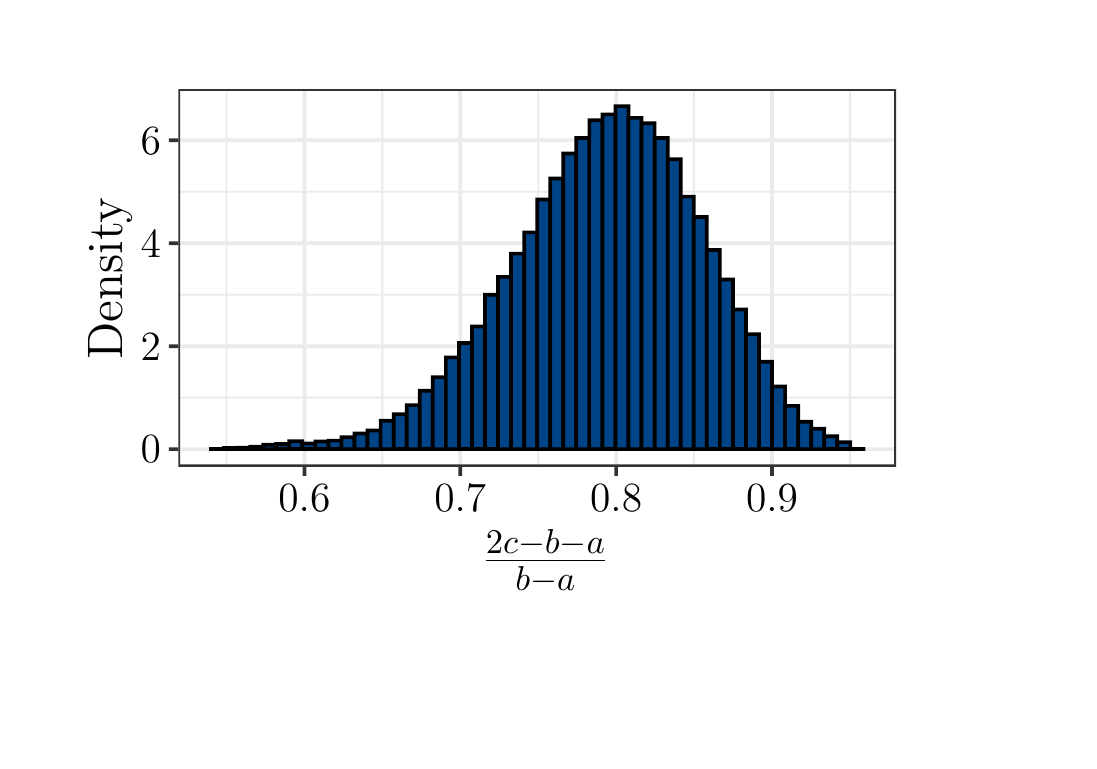}
    \subcaption[]{}
    \label{subfig:SDFT_GT_histogram}
  \end{subfigure}
  \hfill
  \begin{subfigure}[t]{.485\textwidth}
    \centering
    \includegraphics[trim = 17 47 57 19, clip]{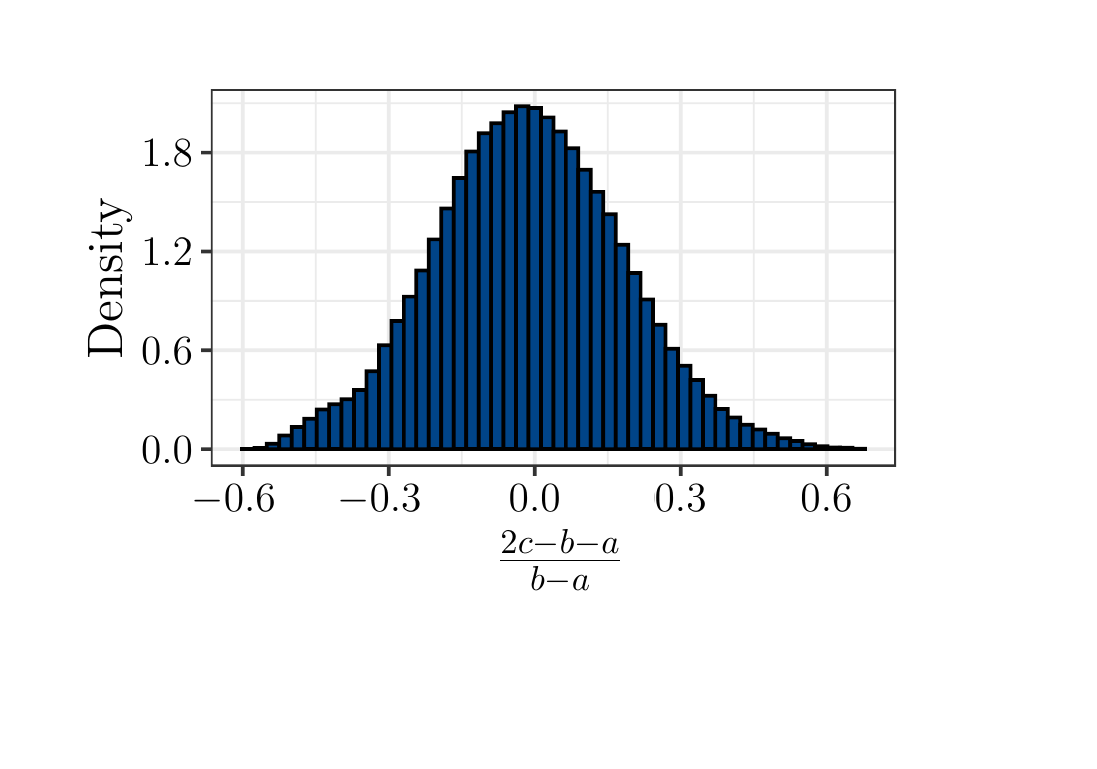}
    \subcaption[]{}
    \label{subfig:CDET_GT_histogram}
  \end{subfigure}
    \caption{Histograms of $\frac{2c - b - a}{b - a}$ for (a) the SDFT data and (b) the CDET data; $\frac{2c - b - a}{b - a}$ ranges between -1 ($c = a$) and 1 ($c = b$) and 0 corresponds to an ST distribution.}
    \label{fig:GT_histograms}
\end{figure}

To investigate possible asymmetry of $f(\lambda_r)$, we applied the GT version of the model within the inference to the SDFT and CDET data. As with the ST version of the model, we produced a chain of 1.5 million samples for the SDFT data and 10.5 million for the CDET data. The posteriors and contour plots obtained are plotted in Figures \ref{fig:SDFT_GT_correlogram} and \ref{fig:CDET_GT_correlogram}. To evaluate the skewness of the GT distributions, we plotted histograms of the quantity $\frac{2c - b - a}{b - a}$ in Figure \ref{fig:GT_histograms}. This quantity can take values between -1 and 1 and is equal to 0 for an ST distribution.

Contour plots and marginals of the posterior are shown in Figure \ref{fig:SDFT_GT_correlogram}. The estimated values of the constitutive parameters are again realistic, with the 95\% confidence interval of $\phi E$ being 823-833 MPa. We note, by comparing with the ST case, that the inclusion of the additional parameter $c$ has slightly increased the predicted values of $\phi E$ and decreased those of $(1-\phi)\mu$. 

In Figure \ref{fig:CDET_GT_correlogram}, we can see that the contour plots and histograms are not as smooth as for the SDFT data, indicating that even more samples might be required to achieve a high degree of convergence. The contour lines also possess a more complex shape than we see for the SDFT data. These features are also present when using the ST model (Figures \ref{fig:SDFT_correlogram} and \ref{fig:CDET_correlogram}). In the GT model fit, a 95\% confidence interval for $\phi E$ is 1340-1380 MPa, which is the same as the ST case to three significant figures. Compared to Figure \ref{fig:CDET_correlogram}, the posteriors for $a$ and $b$ are also similar. 

Finally, Figure \ref{fig:GT_histograms} demonstrates that the value of $c$, the modal fibril length, is predicted to be significantly closer to $b$ than $a$ for the SDFT data, suggesting that this tendon may have a skewed fibril length distribution. In contrast, for the CDET data, $c$ covers a range of values but is most commonly found near the middle of $a$ and $b$, indicating that the distribution is close to symmetric in this case.

\section{Discussion}

In this paper, we developed a new model of soft tissue mechanical behaviour that only contains microstructurally relevant parameters and used a Bayesian MCMC framework to determine the likely range of their values. Our approach was similar to that of Akintunde \textit{et al.} \cite{akintunde2018evaluation}, who used a similar framework to compare the ability of three existing SEFs to fit age-dependent, murine tendon stress-strain data. These SEFs, which were derived by Gasser, Ogden and Holzapfel (GOH) \cite{gasser2006hyperelastic}, Freed and Rajagopal (FR) \cite{freed2016promising} and Shearer (SHR) \cite{shearer2015new} take different approaches to modelling soft tissue fibres. The GOH model accounts for fibril recruitment phenomenologically using an exponential function, whereas the FR model uses an implicit elasticity approach with a phenomenologically chosen implicit energy function. Neither of these models can be used to predict fibril length distributions. The SHR model, on the other hand, makes the specific assumption that fibril length varies radially within a fascicle, with fibrils at the centre of the fascicle being the shortest, and those on its periphery being the longest. Whilst there is evidence that this assumption holds for some tendons \cite{kastelic1978multicomposite}, it is likely not valid for all, or for other biological soft tissues. This was one of the motivations for developing the new constitutive model presented above, which is more general. When fitted to experimental murine and equine tendon data using non-linear optimisation, the new SEF provides a closer fit than the SHR model to all four data sets studied and a closer fit than the HGO model to three out of the four data sets, only narrowly providing a worse relative fit to the t6b data set. 

We also implemented an adaptive RWM algorithm to characterise posterior probability distributions to quantify the uncertainty in the values of the fitting parameters. This algorithm samples effectively when fitting to both synthetic and high-resolution experimental data. Furthermore, it samples parameter vectors that provide a close fit to the data, with 95\% credible intervals for the important physical parameter $\phi E$ containing realistic values when compared with existing estimates of the parameters $\phi$ and $E$. It is intriguing that the GT model predicted that the SDFT has a skewed fibril length distribution, but that the CDET's is likely symmetric. As the SDFT and CDET are archetypal examples of energy-storing and positional tendons, respectively, this may point to a potential difference in the collagen fibril arrangements of these two tendon types. Future work could examine the fit of the GT model in the RWM algorithm to a large number of data sets to determine whether there are indeed differences in the skewness of the fibril length distributions of energy-storing and positional tendons, or whether this result is just an example of inter-sample variation. We emphasise that the distributions that we arrive at here only describe the samples used to create the data, and that they do not reflect the natural variation across all tendons.

As the model is pseudoelastic, the Young's and shear moduli predicted by our model are specific to the strain-rates used in the experiments we fitted. The effective moduli would increase with increasing strain-rate. Our findings suggest that $\phi E$ differs from sample to sample. Consequently, either the collagen volume fraction varies between the samples we fitted, or there may not be a universal collagen fibril Young's modulus. In particular, it may be wrong to assign a Young's modulus to collagen on the fibrillar level as molecular differences may cause some fibrils to be stiffer than others. If so, our model would need to be modified to allow for variation in the constitutive, as well as the structural, parameters. The Bayesian approach assumes that our model is `correct' in the sense that it incorporates all of the physics necessary to predict the microstructural and constitutive parameters accurately. If a significant feature is absent from the model, there would be inaccuracies in the predicted parameter values; however, the quality of fit our model achieved, along with the agreement of the predicted parameter values with experimental values reported in the literature, together, support the assumption that the most important physical features of tendon deformation under the experimental conditions of our test data are included. A more comprehensive model can always be developed, however. There are phenomena not explicitly considered here, such as chemo-mechanical coupling and swelling due to variation in water-content \cite{lozano2019water} that could affect the parameter values in our model; therefore, our results apply only to the specific conditions that were imposed in the experiments we modelled. Explicit incorporation of such effects could be one way to improve upon our model.

We used the triangular distribution to model the distribution of collagen fibril lengths in tendon. Other tractable SEFs could be derived by modelling fibril lengths with alternative distributions, such as the step distribution, for example, which would also lead to a convenient analytic representation. Additionally, more efficient sampling methods, such as Hamiltonian Monte Carlo, which uses derivatives of the log-posterior with respect to model parameters to propose parameter vectors in areas of high posterior probability, could be used instead. We have studied tendons, which possess more strongly aligned collagen than tissues such as skin, where fibrils are generally splayed. By applying our model, and the Bayesian approach used here, to other soft tissues, we could quantify uncertainty in a broader range of scenarios. A plausible test of the model and its assumptions would be to fit multiple data sets simultaneously, enforcing the constitutive parameters to be the same between the data sets and varying the structural parameters only. The posterior probability distributions could then quantify the inter-sample variation in the constitutive parameters of a particular tissue in any given species.

\section*{Author Contributions}
JH implemented the mathematics in Mathematica and R (see Supplementary Material) and wrote the initial draft of the paper. SLC supervised the implementation of the Bayesian MCMC methodology. WJP and TS supervised the implementation of the solid mechanics. TS derived the new strain energy function. All authors contributed to editing and revising the manuscript.

\section*{Acknowledgements}
We thank the reviewers for their helpful and constructive comments.
 
\section*{Data Accessibility}
The paper makes use of two datasets, one collected by Goh \textit{et al.} \cite{goh2008ageing, goh2012bimodal} and one collected by Thorpe \textit{et al.} \cite{thorpe2012specialization}.

Goh \textit{et al.} \cite{goh2008ageing, goh2012bimodal}: The dataset is available at https://figshare.com/collections/Ageing\_tendon\_collection/3938821. The data can be processed in MATLAB using the code available at https://figshare.com/articles/dataset/The\_code\_for\\
\_generating\_and\_processing\_the\_dataset\_for\_load-displacement\_and\_stress-strain/5640649.

Thorpe \textit{et al.} \cite{thorpe2012specialization}: The dataset is available at https://qmro.qmul.ac.uk/xmlui/handle/123456789/13395.

\section*{Funding Statement}
JH is grateful to the Department of Mathematics, University of Manchester for PhD funding. SLC is grateful to the Alan Turing Institute for a Turing fellowship. Parnell is grateful to the Engineering and Physical Sciences Research Council (EPSRC) for funding via grant EP/S019804/1.

\begin{appendix}
\section{The piecewise constants $A(I_4)$, $B(I_4)$, $C(I_4)$, $D(I_4)$ and $G(I_4)$}
The values of the piecewise constants for the general triangular distribution are
\begin{IEEEeqnarray}{rCl}
	A(I_4) &=& \begin{cases} 
    	        0, & I_4 < a^2, \\
    		    -\frac{a^2}{(b-a)(c-a)}, & a^2\leqslant I_4\leqslant c^2, \\
    		    \frac{c^2}{(c-a)(b-c)} - \frac{a^2}{(b-a)(c-a)}, &  c^2 < I_4\leqslant b^2, \\
    		     -1, & I_4>b^2,
    		  \end{cases}, \\
	B(I_4) &=& \begin{cases} 
    	        0, & I_4<a^2, \\
    		    \frac{2a\log a}{(b-a)(c-a)}, & a^2\leqslant I_4\leqslant c^2, \\
    			\frac{2a \log a}{(b-a)(c-a)} - \frac{2c \log c}{(c-a)(b-c)}, & c^2 < I_4\leqslant b^2, \\
    			\frac{2a \log a}{(b-a)(c-a)} + \frac{2b \log b}{(b-a)(b-c)} - \frac{2c \log c}{(c-a)(b-c)}, & I_4>b^2, 
    		   \end{cases} \\ 
	C(I_4) &=& \begin{cases} 
    	        0, & I_4<a^2, \\
    			\frac{1}{(b-a)(c-a)}, & a^2\leqslant I_4\leqslant c^2, \\
    			-\frac{1}{(b-a)(b-c)}, & c^2 < I_4\leqslant b^2, \\
    			0, & I_4>b^2, 
    		   \end{cases} \\
    D(I_4) &=& \begin{cases} 
                0, & I_4<a^2, \\
    			-\frac{2a}{(b-a)(c-a)}, & a^2\leqslant I_4\leqslant c^2, \\
    			\frac{2b}{(b-a)(b-c)}, & c^2 < I_4\leqslant b^2, \\
    			0, & I_4>b^2, 
    		   \end{cases}
\end{IEEEeqnarray}
\begin{equation}
	G(I_4) = \begin{cases} 
	        0, & I_4<a^2, \\
	        \frac{a^2\log a}{(b-a)(c-a)} - \frac{5a^2}{2(b-a)(c-a)}, & a^2\leqslant I_4\leqslant c^2, \\
	        \frac{2a^2\log a}{(b-a)(c-a)} - \frac{c^2\log c}{(c-a)(b-c)} - \frac{5a^2}{2(b-a)(c-a)} + \frac{5c^2}{2(b-c)(c-a)}, & c^2\leqslant I_4\leqslant b^2, \\
	        \frac{a^2\log a}{(b-a)(c-a)} - \frac{c^2\log c}{(c-a)(b-c)} + \frac{b^2\log b}{(b-c)(b-a)} - \frac{5a^2}{2(b-a)(c-a)} + \frac{5c^2}{2(b-c)(c-a)} - \frac{5b^2}{2(b-a)(c-a)}, & I_4>b^2.
	   \end{cases}
\end{equation}
The corresponding quantities for the symmetric triangular distribution are obtained by setting $c=(a+b)/2$.
\end{appendix}

\bibliographystyle{vancouver}
\bibliography{bibfileTendon}

\newpage
\end{document}